\documentclass[reprint,amsmath,amssymb,aps,superscriptaddress,nofootinbib]{revtex4-2}

\usepackage{graphicx}
\usepackage{dcolumn}
\usepackage{bm}
\usepackage{tikz-feynman}
\usepackage{tikz}
\usepackage{caption}
\usepackage{subcaption}
\usepackage{hyperref}
\usepackage[usestackEOL]{stackengine}

\begin{document}
\strutlongstacks{T}
\newcommand{\St}[1]{\Centerstack{#1}}

\title{The reaction $\pi N \to \omega N$ in a dynamical coupled-channel approach}
\author{Yu-Fei Wang}
\email{yuf.wang@fz-juelich.de}
\affiliation{Institute for Advanced Simulation and J\"ulich Center for Hadron Physics, Forschungszentrum J\"ulich, 
52425 J\"ulich, Germany}

\author{Deborah R{\"o}nchen }
\affiliation{Institute for Advanced Simulation and J\"ulich Center for Hadron Physics, Forschungszentrum J\"ulich, 
52425 J\"ulich, Germany}

\author{Ulf-G.~Mei\ss ner }
\affiliation{Helmholtz-Institut f\"ur Strahlen- und Kernphysik (Theorie) and Bethe Center for Theoretical
Physics,  Universit\"at Bonn, 53115 Bonn, Germany}
\affiliation{Institute for Advanced Simulation and J\"ulich Center for Hadron Physics, Forschungszentrum J\"ulich, 
52425 J\"ulich, Germany}
\affiliation{Tbilisi State University, 0186 Tbilisi, Georgia}

\author{Yu Lu}
\affiliation{School of Physical Sciences, University of Chinese Academy of Sciences,
    100049 Beijing, China}

\author{Chao-Wei Shen}
\affiliation{Institute for Advanced Simulation and J\"ulich Center for Hadron Physics, Forschungszentrum J\"ulich, 
52425 J\"ulich, Germany}

\author{Jia-Jun Wu }
\affiliation{School of Physical Sciences, University of Chinese Academy of Sciences,
    100049 Beijing, China}

\date{\today}

\begin{abstract}
A refined investigation on light flavor meson-baryon scatterings is performed using a dynamical coupled-channel approach, the J\"ulich-Bonn model, that respects unitartiy and analyticity constraints. The channel space of $\pi N$, $\pi \Delta$, $\sigma N$, $\rho N$, $\eta N$, $K \Lambda$ and $K \Sigma$ is extended by adding the $\omega N$ final state. The spectra of $N^*$ and $\Delta$ resonances are extracted in terms of complex poles of the scattering amplitudes, based on the result of a global fit to a worldwide collection of data, in the energy region from the $\pi N$ threshold to center-of-mass energy $z=2.3$~GeV. A negative value of the $\omega N$ elastic spin-averaged scattering length is extracted, questioning the existence of bound states of the $\omega$ meson in the nuclear matter. 
\end{abstract}
\maketitle
\section{Introduction}
The scattering of light mesons and baryons, starting from the pion-nucleon channel, has been a topic of interest for many decades and continues to be so. On the one hand, such reactions reveal crucial information to obtain a deeper understanding of the strong interaction, on the other hand, the output from the study of such scattering processes can be used as input for further research, such as nuclear structure or nuclear astrophysics. On the fundamental level, Quantum Chromodynamics (QCD) governs the hadronic interactions. However, due to color confinement, QCD cannot be applied directly when the energy is not high enough. To deal with non-perturbative problems at low energies, chiral perturbation theory~\cite{Weinberg:1966fm,Weinberg:1968de,Coleman:1969sm,Callan:1969sn,Jenkins:1990jv,Meissner:1993ah,Leutwyler:1993iq,Ecker:1994gg,Bernard:1995dp,Scherer:2002tk,Bernard:2006gx} was developed, based on the spontaneously broken chiral symmetry of light flavor quarks and regarding the light hadrons as basic degrees of freedom, with the pseudoscalar mesons being the quasi Goldstone bosons. By pertubative expansions in the meson masses and the momenta, the near-threshold observables can be well reproduced, see e.g.~\cite{Bernard:1993fp,Mojzis:1997tu,Fettes:1998ud,Becher:1999he,Fettes:2001cr,Hoferichter:2009gn,Alarcon:2012kn,Chen:2012nx}, and some low-lying resonances e.g. the $\Delta(1232)$ isobar and the Roper resonance can also be treated, see e.g.~\cite{Jenkins:1991es,Hemmert:1997ye,Beane:2002ud,Pascalutsa:2006up,Borasoy:2006fk,Yao:2016vbz,Siemens:2016hdi,Gegelia:2016xcw,Siemens:2016jwj,Siemens:2017opr}.

Another challenging subject is the intermediate energy region: though experimental observations are abundant, the dynamics becomes more involved as more interaction channels open, and theoretical approaches are further complicated by the appearance of a large amount of resonances. In this region, the effective field theory description breaks down while perturbative QCD is not yet applicable. Extracting the resonances from experimental data is, thus, a fundamental task of great importance, which cannot be trivially accomplished, since most of the states do not exhibit a typical Breit-Wigner behavior. To attack this problem, efforts were made by partial wave analyses~\cite{Arndt:2006bf,Workman:2008iv} restrained to $\pi N$ elastic scattering, resulting in most of the resonances known today. However, due to the fact that some states may not couple strongly to $\pi N$~\cite{Koniuk:1979vw}, coupled-channel anaylses are also necessary. Straightforwardly, the phenomenological inputs can be unitarized or re-summed~\cite{Bruns:2010sv,Lutz:2001mi,Penner:2001fu,Penner:2002ma,Penner:2002md,Shklyar:2004ba}, reproducing unitary amplitudes with complex poles on the unphysical Riemann sheet as the resonances. Furthermore, though complicated, coupled-channel models evolving under dynamical scattering equations~\cite{Mueller:1990uxr,Lohse:1990ew,Pearce:1991xt,Schutz:1994wp,Schutz:1994ue,Schutz:1998jx,Krehl:1999km,Gasparyan:2003fp,Matsuyama:2006rp,Paris:2008ig,Doring:2009yv,Doring:2010ap,Ronchen:2012eg,Khemchandani:2013nma,Ronchen:2015vfa,Mai:2021vsw,Mai:2021aui} do not only keep the unitarity but also globally lead to a better analytical behavior with fewer model artefacts in the complex energy plane. 

The approach applied in this paper is called the J{\"u}lich-Bonn (J\"uBo) model, which has experienced three decades of development. Its core is a Lippmann--Schwinger-like equation, taking tree-level diagrams and correlated two-pion exchange as the kernel. Based on early studies on meson-meson and $K^- N$ interactions~\cite{Mueller:1990uxr,Lohse:1990ew,Pearce:1991xt}, the approach was first applied to $\pi N$ elastic scattering in Refs.~\cite{Schutz:1994wp,Schutz:1994ue}. Its early coupled-channel extensions (including $\pi\pi N$ and $\eta N$) can be found in Refs.~\cite{Schutz:1998jx,Krehl:1999km,Gasparyan:2003fp}. In Ref.~\cite{Doring:2009yv} the analytical structures of the amplitudes obtained by this model are systematically studied, enabling the extraction of resonances as poles in a modern way. In Refs.~\cite{Doring:2010ap,Ronchen:2012eg}, the model has progressed to the zero-strangeness kaon-hyperon channels $K\Lambda$, $K\Sigma$ and higher partial waves up to $J=9/2$, with the center-of-mass energy up to $2.3$~GeV. A major step towards a reliable determination of the light baryon spectrum based on an extensive high-quality data base with the extension of the framework to meson photoproduction was achieved in Refs.~\cite{Ronchen:2014cna,Ronchen:2015vfa,Ronchen:2018ury}. Moreover, in Refs.~\cite{Mai:2021vsw,Mai:2021aui} the approach was adapted to include virtual photons and was applied in the first-ever coupled-channel study of pion and eta electroproduction data. Besides the light meson-baryon scattering and photoproduction processes, the $P_c$ states in the hidden-charm sector were also studied with the J\"uBo model~\cite{Shen:2017ayv,Wang:2022oof}. 

To summarize, the hadronic J{\"u}lich-Bonn model is one of the theoretically best founded tools for studying the spectrum of the light-flavored baryon resonances. So far, the maximum energy considered in the model is $2.3$~GeV but certain channels like $\omega N$ have not been considered. To further refine the model and to enrich the resulting physics, here we extend this model by adding the $\omega N$ channel. Furthermore, the $\omega$ can be considered as the first vector meson regarded as a stable particle in this approach. 

Reaching beyond the completion of our model, the $\omega N$ channel is of interest for several reasons. It has been found that the spontaneous broken chiral symmetry of QCD shows a tendency of restoration as the density of nucleons gets larger in the nuclear matter, as signaled e.g. by dropping vector meson masses~\cite{Brown:1991kk,Hatsuda:1991ez,Cohen:1991nk}, although this particular claim is at odds with earlier calculations~\cite{Bernard:1988db}. Much experimental and theoretical work has been done in this field. On the experimental side, the pertinent signals are usually of electromagnetic nature. For a review, see e.g.~\cite{Rapp:2009yu}. Due to the vector meson dominance~\cite{Gell-Mann:1961jim}, the $\gamma N$ interaction is dominated by $\rho N$, $\omega N$ and $\phi N$. The $\rho$ meson is very broad and would not behave like a quasi-particle when it is in-medium, while the direct $\phi NN$ coupling is suppressed due to the nearly pure $s\bar{s}$ component of $\phi$. In this respect, $\omega N$ can be considered as the most important channel. Additionally, the details of $\omega N$ interaction are essential for understanding the equation of state of the neutron stars~\cite{Shen:1998gq}. 

According to the low-density theorem~\cite{Dover:1971hr}, the additional in-medium self-energy of the $\omega$ is proportional to the elastic scattering amplitude of $\omega N$. Therefore, the elastic $\omega N$ amplitude provides crucial information about the in-medium $\omega$. This holds especially for the real part of the spin-averaged $\omega N$ scattering length, which indicates whether or not the $\omega$ can form bound states in the medium. However, this scattering length cannot be observed directly by experiments. On the theoretical side, even the sign of the spin-averaged $\omega N$ scattering length is still an open question. Results based on QCD sum rules~\cite{Koike:1996ga,Klingl:1997kf,Klingl:1998zj} support an attractive force in the $\omega N$ system from the scattering length, contradicting other analyses~\cite{Lutz:2001mi,Penner:2001fu,Penner:2002ma,Penner:2002md,Shklyar:2004ba,Muehlich:2006nn,Paris:2008ig,Strakovsky:2014wja,Metag:2017yuh,Ishikawa:2019rvz}. Comprehensive models like the current one, which extract the scattering length constrained by a global fit to all possible data sets, are needed to clarify this issue. 

This paper is organized as the follows. In Sect.~\ref{sec:th} we outline the underlying theoretical framework of the model, especially the structure of the scattering equation. All the important information about the fit, like the numerical details and the fit to the data of $\omega N$ channel, as well as the two different solutions to estimate the uncertainties, are displayed and discussed in Sect.~\ref{sec:fit}. Sect.~\ref{sec:res} contains the results concerning the hadron spectrum (baryon resonances) and the scattering lengths, with discussions on the resulting physics. Sect.~\ref{sec:con} presents the conclusions of this work and some perspectives for further studies. The expressions for the observables in the $\omega N$ channel are shown in Appendix~\ref{app:expr}. The partial wave amplitudes of $\pi N\to\omega N$ interaction are given in Appendix~\ref{app:oNpw}. Tables containing the coupling strengths of the resonances to the effective three-body channels ($\pi\Delta$, $\sigma N$ and $\rho N$) are given in Appendix~\ref{app:3bodycoup}. Fit results for channels other than $\omega N$ can be found on the website~\cite{JuBo}. Further details of the theoretical framework are summarized in the supplemental material~\cite{SM}. 

\section{Theoretical framework}\label{sec:th}

\begin{figure*}[t!]
	\centering
	\includegraphics[width=0.8\textwidth]{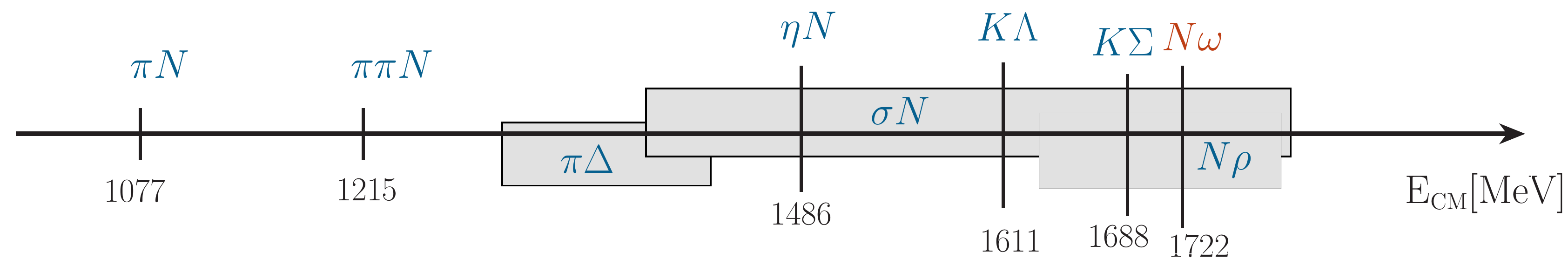}
	\caption{Thresholds of the scattering channels currently considered in the J\"{u}Bo model as function of the center-of-mass energy. }
	\label{fig:channelthrs}
\end{figure*}

\noindent
In the current study, the $\pi N$, $\pi\pi N$, $\eta N$, $K \Lambda$, $K \Sigma$ and $\omega N$
channels are considered. The $\pi\pi N$ system is simulated by three effective channels,
the $\pi \Delta$, $\sigma N$ and $\rho N$. The threshold of each channel is shown in
Fig.~\ref{fig:channelthrs}. Note that the width of the $\omega$ meson is much smaller than its mass,
and is thus not considered in this work. 

The master formula of this model is the following scattering equation: 
\begin{equation}\label{scequ}
\begin{split}
  &T_{\mu\nu}(p'',p',z)=V_{\mu\nu}(p'',p',z)\\
  &+\sum_{\kappa}\int_0^\infty p^2 dp V_{\mu\kappa}(p'',p,z)G_{\kappa}(p,z)T_{\kappa\nu}(p,p',z)\ ,
\end{split}
\end{equation}
where $T$ denotes the scattering amplitude, $V$ denotes the interaction kernel (the potential), $p'$
and $p''$ are the three-momenta of the initial and final states in the center-of-mass frame, respectively,
$z$ is the center-of-mass energy, which is related to $p'$ and $p''$ by the on-shell conditions
when calculating physical observables. Further, $\mu,\nu$ and $\kappa$ are the channel labels
denoting the meson-baryon system with specific isospin ($I$), angular momentum ($J$, up to $9/2$),
spin ($S$) and orbital angular momentum $L$. $G_{\kappa}(p,z)$ is the propagator of the
intermediate channel: 
\begin{widetext}
\begin{equation}\label{Gdef}
	G_{\kappa}(z,p)=
	\begin{cases}
		(z-E_\kappa-\omega_\kappa+i0^+)^{-1}& (\text{if}\ \kappa\ \text{is a two-body channel})\ ,\\
	  \big[z-E_\kappa-\omega_\kappa-\Sigma_\kappa(z,p)+i0^+\big]^{-1}&
          (\text{if}\ \kappa\ \text{is an effective channel})\ .
	\end{cases}
\end{equation}
\end{widetext}
Further, $E_\kappa,\,\omega_\kappa$ denote the energies of the baryon and the meson in channel $\kappa$,
respectively, with a relativistic dispersion relation, e.g. $E_\kappa=\sqrt{p^2+M_\kappa^2}$.
For the quasi three-body channels, $\Sigma_\kappa$ is the self-energy function of the unstable
particle ($\Delta,\sigma,\rho$). Note that such kind of propagators are derived from (old-fashioned)
time-ordered perturbation theory (TOPT)~\cite{schweber1964} rather than the modern covariant convention.
In combination with a partial-wave projection, the application of TOPT reduces the fully relativistic
Bethe-Salpeter equation~\cite{sterman1993} to the one-dimensional integral of Eq.~\eqref{scequ}.
Accordingly, the potential $V$ is also constructed using TOPT. 

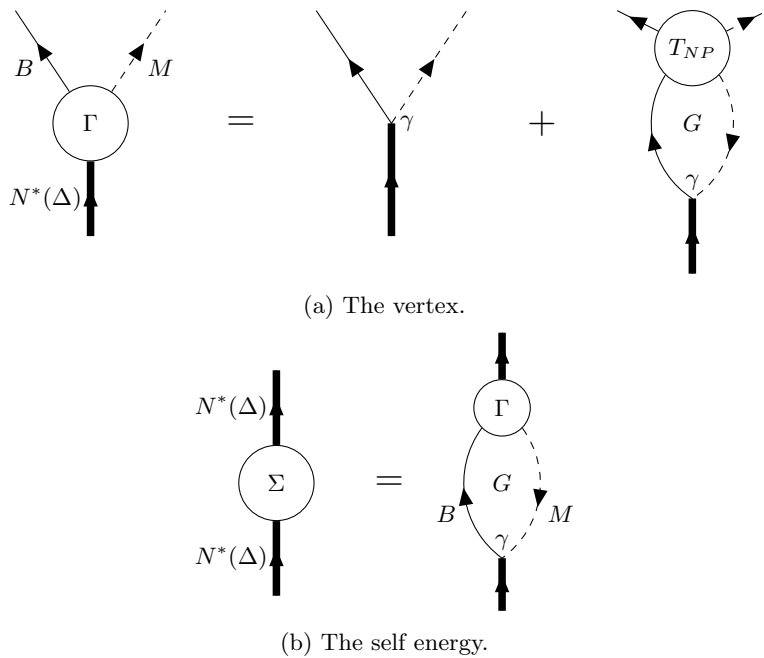
\begin{figure*}[t!]
	\centering
	\begin{subfigure}[b]{0.6\textwidth}
		\centering
		\begin{tikzpicture}
		\begin{feynman}
		\vertex[large,blob,fill=none] (m1) at (0,0) {$\Gamma$};
		\vertex (a1) at ( 0,-1.5);
		\vertex (b1) at (-1, 1.5);
		\vertex (c1) at ( 1, 1.5);
		\node at (2,0) {\Large{$=$}};
		\vertex (m2) at (4,0);
		\node[right] at (4,0) {$\gamma$};
        \vertex (a2) at ( 4,-1.5);
        \vertex (b2) at ( 3, 1.5);
        \vertex (c2) at ( 5, 1.5);
        \node at (6,0) {\Large{$+$}};
        \vertex (a3) at (8,-2);
        \vertex (p) at (8,-1);
        \node[above] at (8,-1) {$\gamma$};
        \node at (8,0) {$G$};
        \vertex[large,blob,fill=none] (q) at (8, 1) {$T_{NP}$};
        \vertex (b3) at ( 7, 1.5);
        \vertex (c3) at ( 9, 1.5);
		\diagram* {
			(a1) -- [fermion,edge label=\(N^*(\Delta)\)] (m1),
			(m1) -- [fermion,edge label=\(B\)] (b1),
			(m1) -- [charged scalar,edge label'=\(M\)] (c1),
			(a2) -- [fermion] (m2),
			(m2) -- [fermion] (b2),
			(m2) -- [charged scalar] (c2),
			(a3) -- [fermion] (p),
			(p) -- [fermion, quarter left] (q) -- [charged scalar, quarter left] (p),
			(q) -- [fermion] (b3),
			(q) -- [charged scalar] (c3),
			
		};
		\end{feynman}
		\draw [line width=1mm] (a1) -- (m1);
		\draw [line width=1mm] (a2) -- (m2);
		\draw [line width=1mm] (a3) -- (p);
		\end{tikzpicture}
		\caption{The vertex. }
		\label{fig:ver}
	\end{subfigure}
	\begin{subfigure}[b]{0.6\textwidth}
		\centering
		\begin{tikzpicture}
		\begin{feynman}
		\vertex[large,blob,fill=none] (m1) at (0,0) {$\Sigma$};
		\vertex (a1) at ( 0,-1.5);
		\vertex (b1) at ( 0, 1.5);
		\node at (1.5,0) {\Large{$=$}};
		\vertex (p) at (3,-1);
		\node[above] at (3,-1) {$\gamma$};
		\node at (3, 0) {$G$};
		\vertex (a2) at (3, -1.7);
		\vertex[blob,fill=none] (q) at (3,1) {$\Gamma$};
		\vertex (b2) at (3, 2);
		\diagram* {
			(a1) -- [fermion,edge label=\(N^*(\Delta)\)] (m1),
			(m1) -- [fermion,edge label=\(N^*(\Delta)\)] (b1),
			(a2) -- [fermion] (p), 
			(p) -- [fermion, quarter left, edge label=\(B\)] (q) -- [charged scalar, quarter left,edge label=\(M\)] (p),
			(q) -- [fermion] (b2),
			
		};
		\end{feynman}
		\draw [line width=1mm] (a1) -- (m1);
		\draw [line width=1mm] (m1) -- (b1);
		\draw [line width=1mm] (a2) -- (p);
		\draw [line width=1mm] (q) -- (b2);
		\end{tikzpicture}
		\caption{The self energy. }
		\label{fig:selfe}
	\end{subfigure}
	\caption{Schematic plot of Eq.~\eqref{Tpequdef}. }\label{fig:DSnstar}
\end{figure*}

In principle the potential $V$  in Eq.~\eqref{scequ} can contain any two-particle irreducible term.
To simplify the calculation, the following separation is performed: 
\begin{equation}
	T=T^{NP}+T^{P}\ .
\end{equation}
Specifically, $T^{NP}$ is generated only by potentials from $t$- and $u$-channel exchange and
contact diagrams, denoted as $V^{NP}$: 
\begin{equation}\label{tnpdef}
\begin{split}
  &T_{\mu\nu}^{NP}(p'',p',z)=V_{\mu\nu}^{NP}(p'',p',z)\\
  &+\sum_{\kappa}\int_0^\infty p^2 dpV_{\mu\kappa}^{NP}(p'',p,z)G_{\kappa}(p,z)T_{\kappa\nu}^{NP}(p,p',z)\ ,
\end{split}
\end{equation}
whereas $T^P$ is constructed via 
\begin{equation}\label{Tpequ}
    T_{\mu\nu}^{P}(p'',p',z)=\sum_{i,j}\Gamma_{\mu,i}^a(p'')(D^{-1})_{ij}(z)\Gamma_{\nu,j}^c(p')\ .
\end{equation}
Here, $i,j$ are the indices of the $s$-channel poles, $\Gamma_{\mu,i}^a$ is the dressed vertex
function describing the annihilation of the $i$th resonance to channel $\mu$, and similar
for $\Gamma_{\nu,j}^c$ (creation of the resonance from channel $\nu$). Also, $D^{-1}$ is
the propagator of the $s$-channel resonances related to the self-energies $\Sigma_{ij}$: 
\begin{equation}\label{Tpequdef}
    \begin{split}
    &\Gamma_{\mu,i}^a(p'')=\gamma_{\mu,i}^a(p'')\\
    &+\sum_{\kappa}\int_0^\infty p^2 dp T_{\mu\kappa}^{NP}(p'',p,z)G_\kappa(p,z)\gamma_{\kappa,i}^a(p)\ ,\\
    &\Gamma_{\nu,j}^c(p')=\gamma_{\nu,j}^c(p')\\
    &+\sum_{\kappa}\int_0^\infty p^2 dp \gamma_{\kappa,j}^c(p)G_\kappa(p,z)T_{\kappa\nu}^{NP}(p,p',z)\ ,\\
    &D_{ij}(z)= \delta_{ij}(z-m_i^b)-\Sigma_{ij}(z)\ ,\\
    &\Sigma_{ij}(z)=\sum_{\kappa}\int_0^\infty p^2 dp \gamma_{\kappa,i}^c(p)G_\kappa(p,z)\Gamma_{\kappa,j}^a(p)\ ;
    \end{split}
\end{equation}
where the $\gamma$'s are bare vertices and $m_i^b$ is the bare mass of the $i$th resonance.
Eq.~\eqref{Tpequdef} actually corresponds to the construction of $T^P$ by Schwinger--Dyson-like equations
illustrated in Fig.~\ref{fig:DSnstar}. 
More importantly, $T^{NP}$ can also dynamically generate poles which are not included as genuine
states in the $s$-channel. 

The potentials $V^{NP}$ are constructed based on the leading order Lagrangians respecting SU(3) flavor symmetry,
$C$ and $P$ conservation, as well as derivative couplings of pseudoscalar mesons from chiral
symmetry~\cite{Wess:1967jq,Kroll:1967it,Meissner:1987ge}. Except for the correlated two-pion
exchange described in detail in Refs.~\cite{Schutz:1994wp,Schutz:1994ue}, all the other potentials
are considered at tree level, causing divergences of the integral in Eq.~\eqref{tnpdef}. Hence we
add regulators to make the integrals converge, thereby introducing cut-off parameters to
be determined by the fit. Such regulators can also be understood as phenomenological form
factors, simulating the inner structures of the hadrons. Additionally the integrals in
Eq.~\eqref{Tpequdef} need similar regulators, but since the bare couplings are already
fit parameters, the $s$-channel cut-offs are fixed. 

A more detailed description of the potential can be found in the supplemental material~\cite{SM}.
Note that besides the $\omega N$ channel, to make the dynamics of the $K\Sigma$ system complete,
in this work we add two more diagrams. These are $K\Lambda\to K\Sigma$
and $K\Sigma\to K\Sigma$ reactions with the $t$-channel $a_0$ exchange. 
They are allowed by SU(3) symmetry and other conservation laws,
but were not considered in the previous studies using the J{\"u}Bo model. 

Note that the nucleon itself is an $s$-channel pole, and we always adjust its bare mass and
couplings such that the physical values are recovered, for details see Ref.~\cite{Ronchen:2012eg}.
In addition to the $s$-, $t$-, and $u$-channel diagrams, just as in Ref.~\cite{Ronchen:2015vfa},
we also add phenomenological contact terms to simulate the effects from physics not explicitly
contained in the $s$-, $t$-, and $u$-channel diagrams. 

The observables can be evaluated using $T_{\mu\nu}$. The normalized, dimensionless partial-wave
amplitudes $\tau$ are directly related to the observables: 
\begin{equation}\label{taudef}
	\tau_{\mu\nu}=-\pi\sqrt{\rho_\mu\rho_\nu}T_{\mu\nu}\ ,
\end{equation}
where $\rho$ is a kinematical phase factor, 
\begin{equation}\label{rhodef}
	\rho_\kappa=\frac{p_{\kappa}}{z}E_\kappa\omega_\kappa\ ,
\end{equation}
and $p_{\kappa}$ is the corresponding three-momentum. The elements of the unitary scattering
$S$-matrix are also written in terms of $\tau$: 
\begin{equation}\label{staurel}
	S_{\mu\nu}=\delta_{\mu\nu}+2i\tau_{\mu\nu}\ .
\end{equation} 

\section{Fits}\label{sec:fit}
\subsection{Database and numerical details}
For the newly included $\omega N$ channel, we fit to the data from the two reactions $\pi^+ n\to
\omega p$ and $\pi^- p\to \omega n$ simultaneously, in addition to all other channels of the
approach. The database for $\omega N$ is summarized in Tab.~\ref{tab:omeNdata}. 
\begin{table*}[t!]
\begin{ruledtabular}
\begin{tabular}{cccc}
Ref. & Observable & Data points & Reaction\\
\hline
Table.~V of Ref.~\cite{Danburg:1970ero} &  XS & 13 & $\pi^+n\to \omega p$\\ 
Fig.~20 of Ref.~\cite{Danburg:1970ero} (Ref.~\cite{Kraemer:1964iv} etc.)&  XS & 10 &  $\pi^+n\to \omega p$ and $\pi^-p\to \omega n$\\ 
Fig.~14 of Ref.~\cite{Binnie:1973hfj} &  FWD & 10 & $\pi^-p\to \omega n$\\ 
Fig.~6 of Ref.~\cite{Keyne:1976tj} &  FWD/BWD & $34$& $\pi^-p\to \omega n$\\ 
Fig.~8 of Ref.~\cite{Keyne:1976tj} &  XS & $23$& $\pi^-p\to \omega n$\\ 
Tab.~1 of Ref.~\cite{Karami:1979ib} &  XS & $8$& $\pi^-p\to \omega n$\\ 
Fig.~3 of Ref.~\cite{Karami:1979ib} &  DXS & $80$ & $\pi^-p\to \omega n$\\ 
Total & - & $178$ & -\\ 
\end{tabular}
\end{ruledtabular}
\caption{The database of $\pi N\to\omega N$. ``XS'' and ``DXS'' refer to total and differential cross sections, respectively. ``FWD(BWD)'' refers to  the differential cross section for the nucleon forward (backward) direction. }
\label{tab:omeNdata}
\end{table*}
The datasets of the channels lower than $\omega N$ remain the same as in Ref.~\cite{Ronchen:2012eg},
see Refs.~\cite{eisler1958bubble,crawford1959charge,baltay1961production,berthelot1961etude,Bertanza:1962pt,Crawford:1962zzb,Yoder:1963zg,Keren:1964ra,carayannopoulos1965,miller1965strange,daronian1966pi,Goussu:1966ps,Dagan:1967pvc,Dahl:1967hix,Doyle:1968zz,Binford:1969ts,Deinet:1969cd,Good:1969rb,VanDyck:1969ay,Kalmus:1970zx,Pan:1970ez,Richards:1970cy,Jones:1971zm,bellamy1972,Knasel:1973ma,Berthon:1974zd,Binnie:1973hfj,Debenham:1975bj,Feltesse:1975nz,Winik:1977mm,Baker:1978qm,Baker:1979aw,Brown:1979ii,Saxon:1979xu,Candlin:1982yv,Bell:1983dm,baldini1988landolt,Candlin:1988pn,Morrisonphd,Arndt:2003if,Kozlenko:2003hu,Prakhov:2005qb,Arndt:2006bf,Bayadilov:2008zz}.
Note that for the $\pi N$ elastic scattering we fit to the energy-dependent GWU/SAID
solution~\cite{Arndt:2006bf}, while the other data points are direct experimental observables.
There are approximately 9000 data points in total in our fit. Moreover, the numbers of fit
parameters are summarized in Tab.~\ref{tab:parmnum}. 
\begin{table}[t!]
\small
\begin{ruledtabular}
\begin{tabular}{cccc}
 & $T^{NP}$ & $T^{P}$ & Total\\
\hline
Newly included & $15$ & $23$ & $38$\\ 
Old &  $64$ & $202$ &  $266$\\ 
Total &  $79$ & $225$ & $304$\\ 
\end{tabular}
\end{ruledtabular}
\caption{The number of fit parameters in this work. See Eqs.~\eqref{tnpdef} and \eqref{Tpequ}
for the separation of $T^{NP}$ and $T^{P}$ (including the phenomenological contact terms).
``Newly included'' means the parameters newly introduced by the $\omega N$ channel
as well as by the two extra $a_0$ exchange potentials in the $K\Sigma$ channel. }
\label{tab:parmnum}
\end{table}

The fits are performed on the JURECA supercomputer at Forschungszentrum J\"ulich~\cite{JUWELS}, where
one node contains $128$ processors. Parallel programming is realized by the Message Passing
Interface (MPI) in Fortran. Every processor takes a value of energy $z$, covering the range from
the $\pi N$ threshold to $2.3$~GeV. At each energy, the scattering equation Eq.~\eqref{tnpdef}
can be solved numerically applying the Haftel-Tabakin scheme~\cite{Haftel:1970zz} by discretizing
the integral via Gaussian quadrature and inverting the $(1-VG)$ matrix algebraically.
The $\chi^2$, measuring the deviation of the theoretical curve to the data points, is
minimized by the MINUIT package~\cite{minuit}. 

As the $\omega N$ data base is quantitatively and qualitatively very limited, we apply weighting
factors in the $\chi^2$ minimization to force the fit to describe details of the data 
that would otherwise be ignored. In addition, the inclusion of the covariance matrices for the
$\pi N\to\pi N$ input from the GWU/SAID solution~\cite{Doring:2016snk} is postponed to
the future. We also note  that some  data sets exhibit questionable uncertainties (e.g. the $\eta N$
differential cross section from Brown et. al.~\cite{Brown:1979ii}), c.f. the
discussion in Ref.~\cite{Ronchen:2012eg}, so that we refrain from giving $\chi^2$ values here.

Further discussions on the systematic uncertainties of our fit results stemming from the applied model
are required. By definition, for a model a systematic error is difficult to determine. We have,
however, tried to obtain a rough estimate of the model-dependence by introducing two different fit
scenarios and comparing their results: 
\begin{itemize}
\item Fit~A: a solution with dynamically generated poles similar to those in Ref.~\cite{Ronchen:2012eg}. 
\item Fit~B: starts from an intermediate stage of fit~A, with an extra narrow (dynamical) pole in
  the $P_{11}$ ($J^P=\frac{1}{2}^+$) wave, with a description of the data of equal quality. 
\end{itemize}

An in-depth uncertainty analysis using, e.g., the least absolute shrinkage and selection
operator (LASSO)~\cite{LASSO} or Bayesion evidence to determine the significance of resonance
states, is planned for the future.

\subsection{Description of the data}
\begin{figure}[htbp]
	\centering
	\begin{subfigure}[b]{0.48\textwidth}
		\centering
		\includegraphics[width=1.0\textwidth]{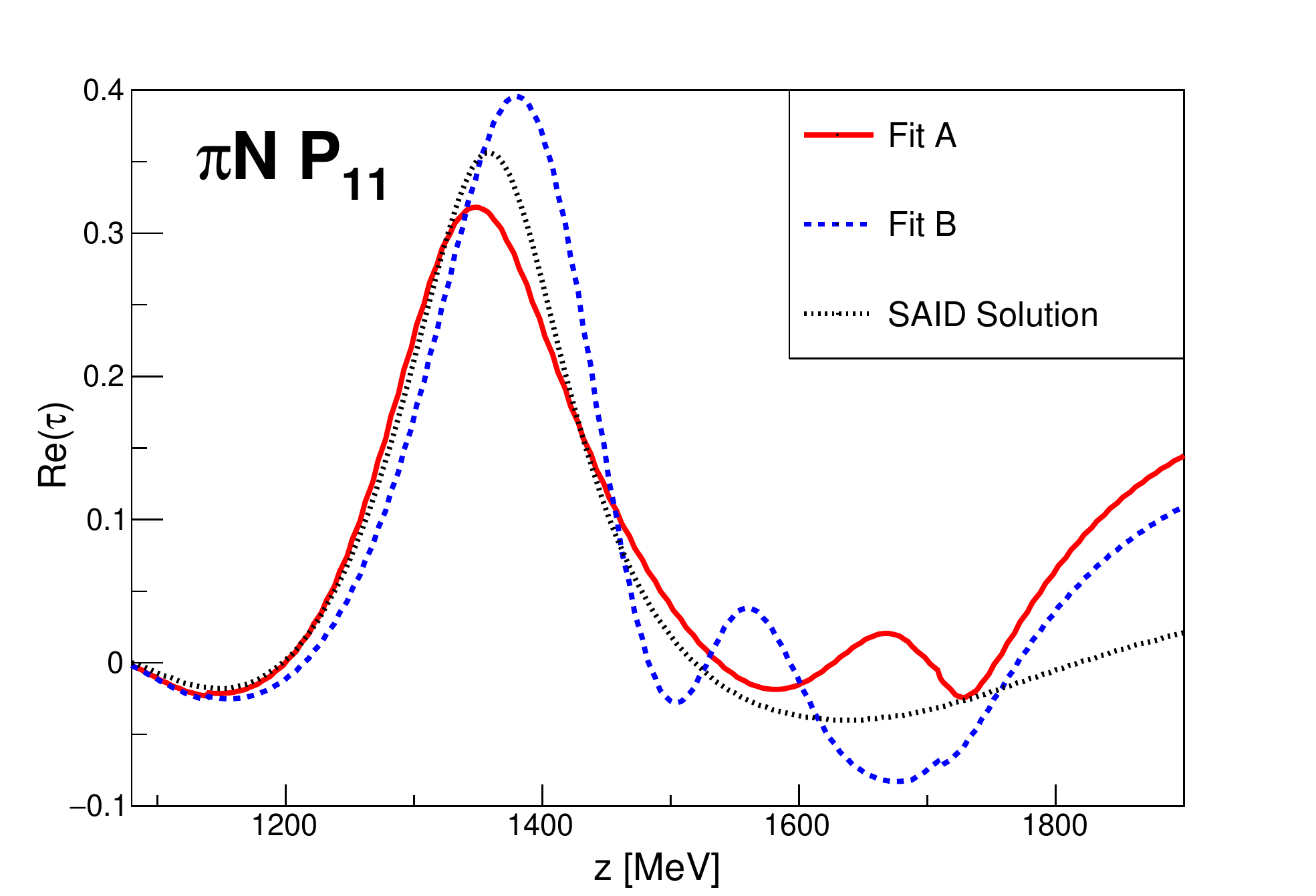}
		\caption{The real part. }
		\label{fig:P11Re}
	\end{subfigure}
	\begin{subfigure}[b]{0.48\textwidth}
		\centering
		\includegraphics[width=1.0\textwidth]{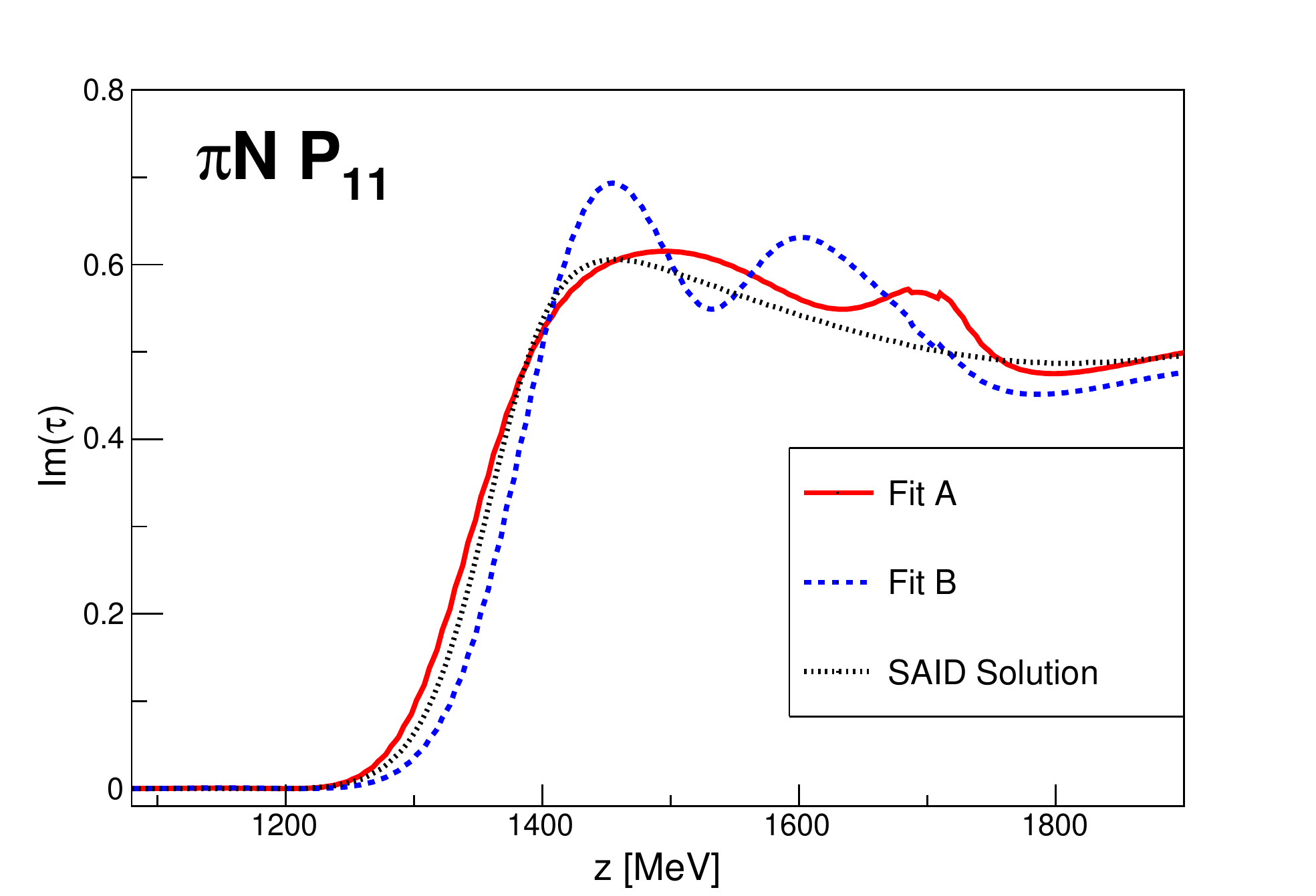}
		\caption{The imaginary part. }
		\label{fig:P11Im}
	\end{subfigure}
	\caption{The comparison of fits~A and B for the $P_{11}$ $\pi N$ elastic $\tau$ amplitude.
          The SAID solution is from Ref.~\cite{Arndt:2006bf} (energy-dependent solution). }
        \label{fig:AvsB}
\end{figure}

As for the description of the data, fit~B is mainly distinguished from fit~A by the $\pi N$ amplitude
of the $P_{11}$ ($J^P=\frac{1}{2}^+$) partial wave, see Fig.~\ref{fig:AvsB}. In both results there is a
(weak) kink when the energy is around $1700$~MeV. This is due to the inclusion of the $N^*(1710)$
resonance, which is absent in
Ref.~\cite{Arndt:2006bf}. Furthermore, fit~B gives another structure when the energy is
around $1500$~MeV,
which, as an effect from a nearby pole, will be discussed later. Despite such a discrepancy,
the global fit qualities are good in both cases.

In the following we only focus on the $\omega N$ channel, since the observables of the other
channels are the same as in Ref.~\cite{Ronchen:2012eg}, and the new fit results are shown
on the website~\cite{JuBo}. The expressions of the differential cross section and total cross section
are summarized in Appendix~\ref{app:expr}. 

\begin{figure}[htpb]
	\centering
	\includegraphics[width=0.55\textwidth]{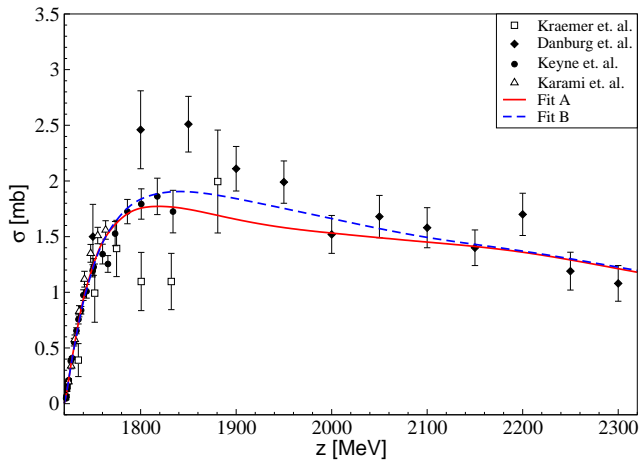}
	\caption{The total cross section for $\pi N\to \omega N$ reaction. The data are from
          Kraemer et al.~\cite{Kraemer:1964iv}, Danburg et al.~\cite{Danburg:1970ero}, Keyne
          et al.~\cite{Keyne:1976tj} and Karami et al.~\cite{Karami:1979ib}. }
	\label{fig:onTXS}
\end{figure}

\begin{figure}[t!]
	\centering
	\begin{subfigure}[b]{0.48\textwidth}
		\centering
		\includegraphics[width=1.0\textwidth]{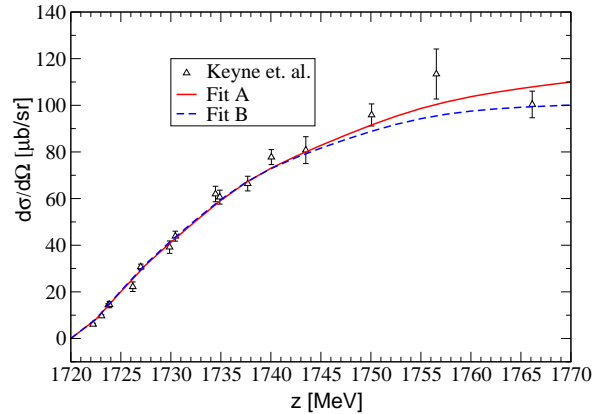}
		\caption{Backward differential cross section. }
		\label{fig:bwd}
	\end{subfigure}
	\begin{subfigure}[b]{0.48\textwidth}
		\centering
		\includegraphics[width=1.0\textwidth]{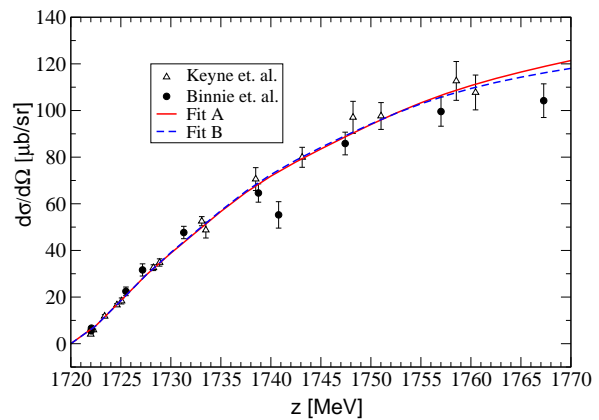}
		\caption{Forward differential cross section. }
		\label{fig:fwd}
	\end{subfigure}
	\caption{The backward and forward differential cross sections of $\pi N\to \omega N$. The
          data are from Binnie et al.~\cite{Binnie:1973hfj} and Keyne et al.~\cite{Keyne:1976tj}.}
        \label{fig:fbwd}
\end{figure}

\begin{figure*}[t!]
	\centering
	\includegraphics[width=0.8\textwidth]{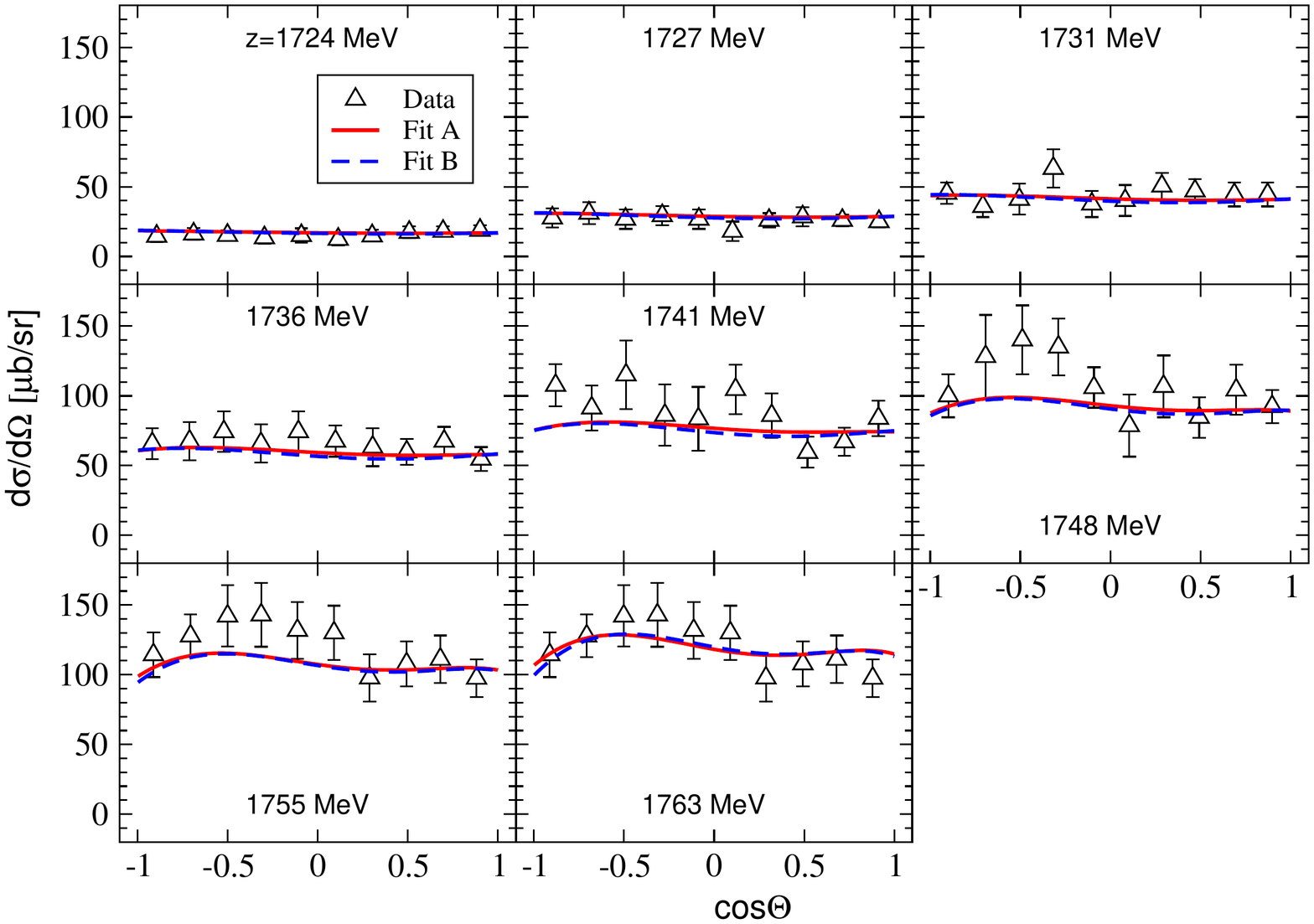}
	\caption{Differential cross section of $\pi N\to \omega N$ at different energies.
          The data are from Ref.~\cite{Karami:1979ib}.}
	\label{fig:onDXS}
\end{figure*}

Both fits~A and B give a quite good description of the $\omega N$ data. The description of the
total cross section is shown in Fig.~\ref{fig:onTXS}. The difference between the two results
becomes visible when $z>1800$~MeV, where the amount of data points decreases. 
As for the differential cross sections, the results for the nucleon backward and forward
directions are displayed in Fig.~\ref{fig:fbwd}. The curves from fits~A and B are quite close
to each other, especially when the energy is near the $\omega N$ threshold. 
The angular distributions of the differential cross sections are also fitted well, see
Fig.~\ref{fig:onDXS}. 

\begin{figure*}[htbp]
	\centering
	\includegraphics[width=0.6\textwidth]{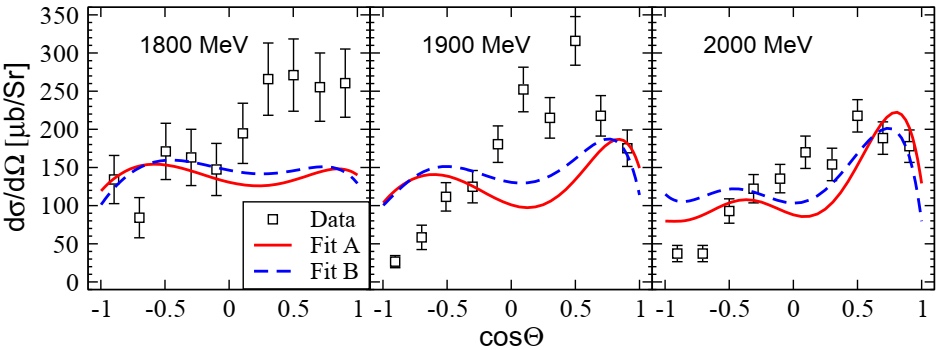}
	\caption{The $\pi N\to \omega N$ differential cross sections
     for the CM energy $1800$ to $2000$ MeV. The data points were extracted from the 
     histograms in Ref.~\cite{Danburg:1970ero} by Ref.~\cite{Penner:2001fu}, 
	 and are not included in the fit. See text for further explanations. }
    \label{fig:histo}
\end{figure*}

In this work we just fit to the original experimental data summarized in Tab.~\ref{tab:omeNdata}. In addition, there are $30$ more points for energies $z=1800,1900$ and $2000$ MeV, which were actually extracted from the histograms of Ref.~\cite{Danburg:1970ero} by Ref.~\cite{Penner:2001fu}. As a comparison, we also plot the differential cross sections at those energies in Fig.~\ref{fig:histo} from our model. It shows that our predictions at these energies are not far away from the data. 

At last, based on those two fit results, the partial wave amplitudes of $\pi N\to \omega N$
reaction are plotted in Appendix~\ref{app:oNpw}. 
\section{Results}\label{sec:res}
\subsection{$N^*$ and $\Delta$ spectrum}
\subsubsection{Pole parameters}
In this model the resonances are extracted by searching for complex poles of the amplitude $\tau$
on the unphysical (i.e. second) Riemann sheet, as described in detail in Ref.~\cite{Doring:2009yv}.
For one resonance pole\footnote{Resonance poles always appear as pairs: when $z_0$ is a pole then
  $z_0^*$ must be another. Here we only discuss the one with the negative imaginary part.},
\begin{equation}
	z_0=M-\frac{\Gamma}{2}i\ ,
\end{equation}
where $M$ is the pole mass and $\Gamma$ the pole width. The leading order Laurent expansion is
parametrized as
\begin{equation}\label{taupole}
	\tilde{\tau}_{\mu\nu}\sim \frac{R_\mu R_\nu}{z_0-z}+\cdots\ ;
\end{equation}
where $\tilde{\tau}$ represents the $\tau$ amplitude of the reaction channels $\nu\to\mu$ on the
unphysical sheet, and $R_\mu$ is the residue belonging to channel $\mu$. In this paper we use
the convention of Partial Data Group (PDG)~\cite{PDG} to measure the coupling strength
by the so-called normalized residue: 
\begin{equation}\label{NRdef}
	NR_\mu \equiv \frac{2R_{\pi N}}{\Gamma}\times R_\mu\ .
\end{equation}
One can define the nominal partial width in terms of $R_\mu$: 
\begin{equation}\label{Gmudef}
	\Gamma_\mu \equiv |\sqrt{2}R_\mu|^2\ .
\end{equation}
The branching ratios can further be defined via 
\begin{equation}\label{BRdef}
	BR_\mu=\frac{\Gamma_\mu}{\Gamma}\  ,
\end{equation}
and the absolute value of the normalized residue is the transition branching ratio: 
\begin{equation}\label{TBRdef}
	BR_{\pi N\to\mu}=|NR_\mu|=\frac{\sqrt{\Gamma_\mu \Gamma_{\pi N}}}{\Gamma}\ .
\end{equation}
Note that only when the resonance is an ideal Breit-Wigner state, then the definitions above
would meet the common understandings of their names, e.g. the ``branching ratios'' are understood as the possibilities for the resonance decaying to the final states, which sum up to $100\%$. Nevertheless most of the $N^*$ and $\Delta$
resonances are usually distorted by complicated coupled-channel dynamics, resulting 
in strong interferences among one another, one has to take those quantities with a grain of salt. 
\subsubsection{Pole positions}
\begin{figure*}[htbp]
	\centering
	\begin{subfigure}[b]{0.8\textwidth}
		\centering
		\includegraphics[width=0.8\textwidth]{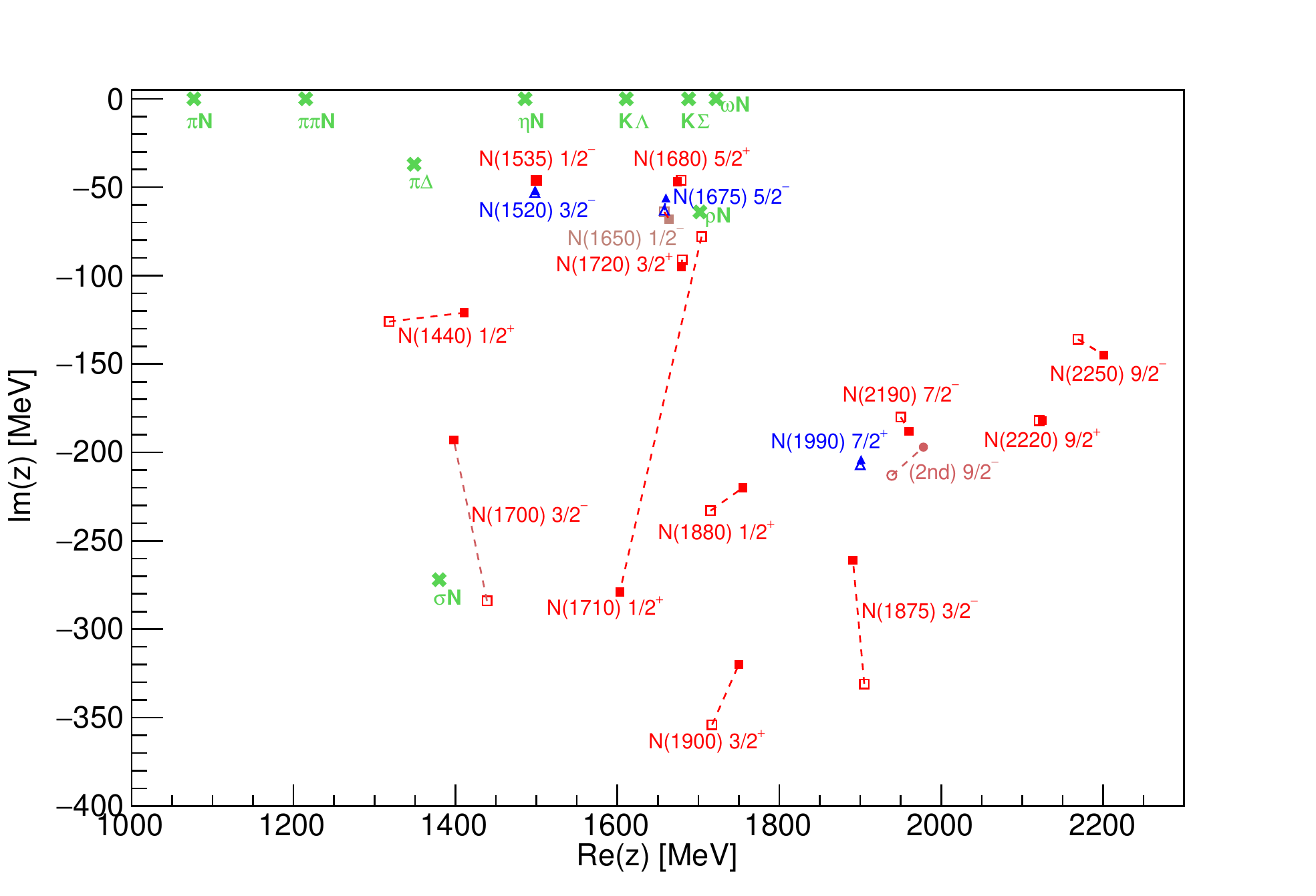}
		\caption{$N^*$ states. }
		\label{fig:Nstloc}
	\end{subfigure}
	\begin{subfigure}[b]{0.8\textwidth}
		\centering
		\includegraphics[width=0.8\textwidth]{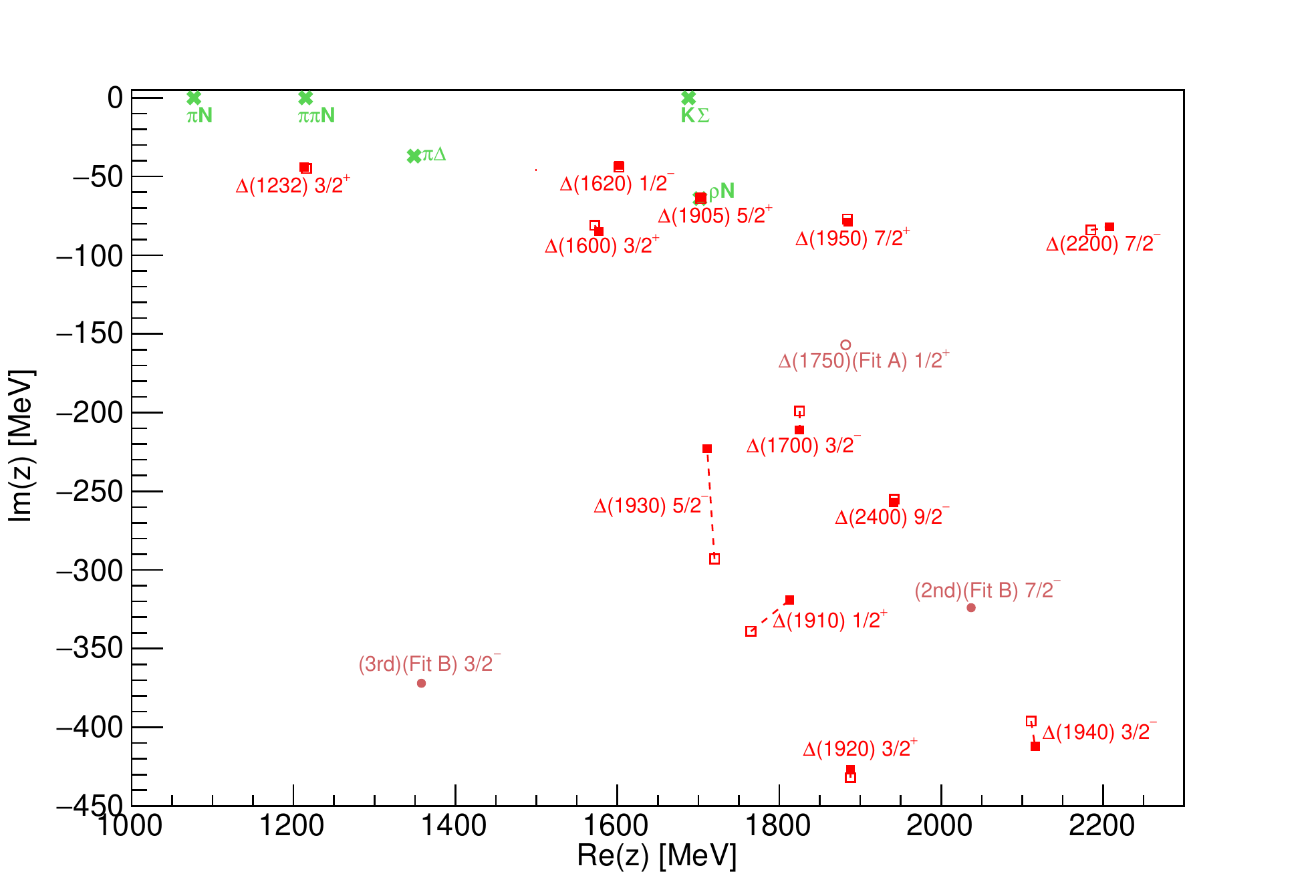}
		\caption{$\Delta$ states. }
		\label{fig:Delloc}
	\end{subfigure}
	\caption{Pole positions from the two fit solutions. Thresholds are labelled by green crosses,
          and the results of fit~A~(B) are denoted by empty (filled) symbols. }\label{fig:poleloc}
\end{figure*}

First, we list the pole positions of both $N^*$ and $\Delta$ resonances within our reach
($\text{Im}(z_0)\leq 450$ MeV) in Tabs.~\ref{tab:poleI1} and \ref{tab:poleI3}, respectively.
Their locations are also plotted in Fig.~\ref{fig:poleloc}. We try to assign the names of most
poles according to PDG~\cite{PDG}\footnote{It must be noted that some of the results published
  in the PDG tables, especially when it comes to the star rating of the corresponding states,
  are debatable.}. Some of the pole positions show a large model-dependence and deviate much from PDG. 
\begin{table*}[t!]
	\small
  \begin{ruledtabular}
	\begin{tabular}{cccc} 
		Resonances & Fit A & Fit B & Estimation of PDG~\cite{PDG} \\
		\hline
		$N(1535)\,\frac{1}{2}^-$ & $1500-46i$ & $1499-46i$ & $1510-65i$ (****)\\ 
		$N(1650)\,\frac{1}{2}^-$ & $1658-64i$ &  $1664-68i$ & $1655-68i$ (****)\\ 
		$N(1440)\,\frac{1}{2}^+$ (NP) & $1318-126i$ & $1411-121i$ & $1370-88i$ (****)\\ 
		$N(1710)\,\frac{1}{2}^+$ & $1704-78i$ &  $1603-279i$ & $1700-60i$ (****)\\ 
		$N(1880)\,\frac{1}{2}^+$ (NP) & $1715-233i$ & $1755-220i$ & $1860-115i$ (***)\\ 
		$N(1720)\,\frac{3}{2}^+$ & $1680-91i$ & $1679-95i$ & $1675-125i$ (****)\\ 
		$N(1900)\,\frac{3}{2}^+$ & $1717-354i$ & $1750-320i$ & $1920-75i$ (****)\\ 
		$N(1520)\,\frac{3}{2}^-$ & $1498-53i$ & $1499-52i$ & $1510-55i$ (****)\\ 
		$N(1700)\,\frac{3}{2}^-$ (NP) & $1439-284i$ & $1398-193i$ & $1700-100i$ (***)\\ 
		$N(1875)\,\frac{3}{2}^-$ (NP) & $1905-331i$ & $1891-261i$ & $1900-80i$ (***)\\ 
		$N(1675)\,\frac{5}{2}^-$ & $1658-63i$ & $1660-56i$ & $1660-68i$ (****)\\ 
		$N(1680)\,\frac{5}{2}^+$ & $1679-46i$ & $1674-47i$ & $1675-60i$ (****)\\ 
		$N(1990)\,\frac{7}{2}^+$ & $1900-207i$ & $1901-204i$ & omitted (**)\\ 
		$N(2190)\,\frac{7}{2}^-$ & $1950-180i$ & $1960-188i$ & $2100-200i$ (****)\\ 
		$N(2250)\,\frac{9}{2}^-$ & $2169-136i$ & $2201-145i$ & $2200-210i$ (****)\\ 
		2nd pole~$\frac{9}{2}^-$ (NP) & $1939-213i$ & $1978-197i$ & $-$\\ 
		$N(2220)\,\frac{9}{2}^+$ & $2121-182i$ & $2125-182i$ & $2170-200i$ (****)\\ 
	\end{tabular}
  \end{ruledtabular}
	\caption{Pole positions of the $N^*$ states (in units of MeVs). The first column also
          exhibits the $J^P$ quantum numbers. ``NP'' means the pole is possibly dynamically
          generated from the $T^{NP}$ part. The significance of the states according to the
          PDG~\cite{PDG} is also shown in the last column.}
	\label{tab:poleI1}
\end{table*}
\begin{table*}[t!]
	\small
  \begin{ruledtabular}
	\begin{tabular}{cccc} 
		Resonances & Fit A & Fit B & Estimation of PDG~\cite{PDG} \\
		\hline
		$\Delta(1620)\,\frac{1}{2}^-$ & $1602-44i$ & $1602-43i$ & $1600-60i$ (****)\\ 
		$\Delta(1750)\,\frac{1}{2}^+$ (NP) & $1882-157i$ &  $-$ & omitted (*)\\
		$\Delta(1910)\,\frac{1}{2}^+$ & $1765-339i$ & $1813-319i$ & $1860-150i$ (****)\\ 
		$\Delta(1232)\,\frac{3}{2}^+$ & $1216-45i$ & $1213-44i$ & $1210-50i$ (****)\\
		$\Delta(1600)\,\frac{3}{2}^+$ (NP) & $1572-81i$ & $1577-85i$ & $1510-135i$ (****)\\
		$\Delta(1920)\,\frac{3}{2}^+$ & $1888-432i$ & $1888-427i$ & $1900-150i$ (***)\\
		$\Delta(1700)\,\frac{3}{2}^-$ & $1825-199i$ & $1825-211i$ & $1665-125i$ (****)\\
		$\Delta(1940)\,\frac{3}{2}^-$ (NP) & $2111-396i$ & $2116-412i$ & $1950-175i$ (**)\\
		3rd pole~$\frac{3}{2}^-$ (NP) & $-$ & $1358-372i$ & $-$\\
		$\Delta(1930)\,\frac{5}{2}^-$ & $1720-293i$ & $1711-223i$ & $1880-140i$ (***)\\
		$\Delta(1905)\,\frac{5}{2}^+$ & $1703-64i$ & $1703-63i$ & $1800-150i$ (****)\\
		$\Delta(1950)\,\frac{7}{2}^+$ & $1884-77i$ & $1885-79i$ & $1880-120i$ (****)\\
		$\Delta(2200)\,\frac{7}{2}^-$ & $2185-84i$ & $2208-82i$ & $2100-170i$ (***)\\
		2nd pole~$\frac{7}{2}^-$ (NP) & $-$ & $2037-324i$ & $-$\\ 
		$\Delta(2400)\,\frac{9}{2}^-$ & $1942-255i$ & $1941-257i$ & omitted (**)\\
	\end{tabular}
  \end{ruledtabular}
	\caption{Pole positions of the $\Delta$ states (in units of MeVs). The first column also
          exhibits the $J^P$ quantum numbers. ``NP'' means the pole is possibly dynamically generated
          from the $T^{NP}$ part. The significance of the states according to the PDG~\cite{PDG} is
          also shown in the last column.}
	\label{tab:poleI3}
\end{table*}

As for the $N^*$ states, some well-established states are rather stable in different fits or even
different studies (see e.g. Ref.~\cite{Ronchen:2012eg}). These are the $N(1535)\,\frac{1}{2}^-$,
$N(1650)\,\frac{1}{2}^-$, $N(1720)\,\frac{3}{2}^+$, $N(1520)\,\frac{3}{2}^-$, $N(1675)\,\frac{5}{2}^-$
and $N(1680)\,\frac{5}{2}^+$. It is always hard to determine the high-spin states, but some
$J\geq 7/2$ states are stable in this study. In this model the lineshape of the $P_{11}\,\pi N$ amplitude
is dominated by the $T^{NP}$ term, with $N(1710)\,\frac{1}{2}^+$ being the only genuine $s$-channel
state. This feature has already been found earlier, see e.g. Ref.~\cite{Krehl:1999km}. It is
understandable that the $\frac{1}{2}^+$ poles in fit~A derivate from those in fit~B, since
the two solutions are distinguished by crucial $T^{NP}$ parameters. In the $P_{13}$ partial wave,
the $N(1900)3/2^+$ moved far into the complex plain and is much broader than in J\"uBo studies
including photoproduction reactions~\cite{Ronchen:2018ury,Ronchen:2022}. This supports the
observation in Refs.~\cite{Ronchen:2018ury,Ronchen:2022} and many other studies by different
analysis groups that this state is especially important in kaon photoproduction, which is not
included in the present work, and hard to determine from purely hadronic reactions. Moreover, the
$N(1700)\,\frac{3}{2}^-$ pole in this work is lower and broader than the estimation in PDG~\cite{PDG}.
Last, we emphasize that the $N(1895)\,\frac{1}{2}^-$, from the coupled-channel analyses of
photonproduction in Ref.~\cite{Anisovich:2011fc}, is of 4-star significance in PDG~\cite{PDG},
but until now it {\it has never been needed in the J\"uBo model} to obtain a good description of the data. 

As already mentioned, the different lineshape of fit~B in the $P_{11}$ partial wave, see
Fig.~\ref{fig:AvsB}, stems from an extra narrow resonance at $z_0=1585-35i$~MeV, with strong
couplings to all the channels. However, in fit~A this pole moves to $1490-245i$ MeV. Indications
for similar broad resonance are observed in a recent update of J\"uBo approach including
$K\Sigma$ photoproduction data~\cite{Ronchen:2022}. 

\begin{figure}[t!]
	\centering
	\begin{subfigure}[b]{0.48\textwidth}
		\centering
		\includegraphics[width=1.0\textwidth]{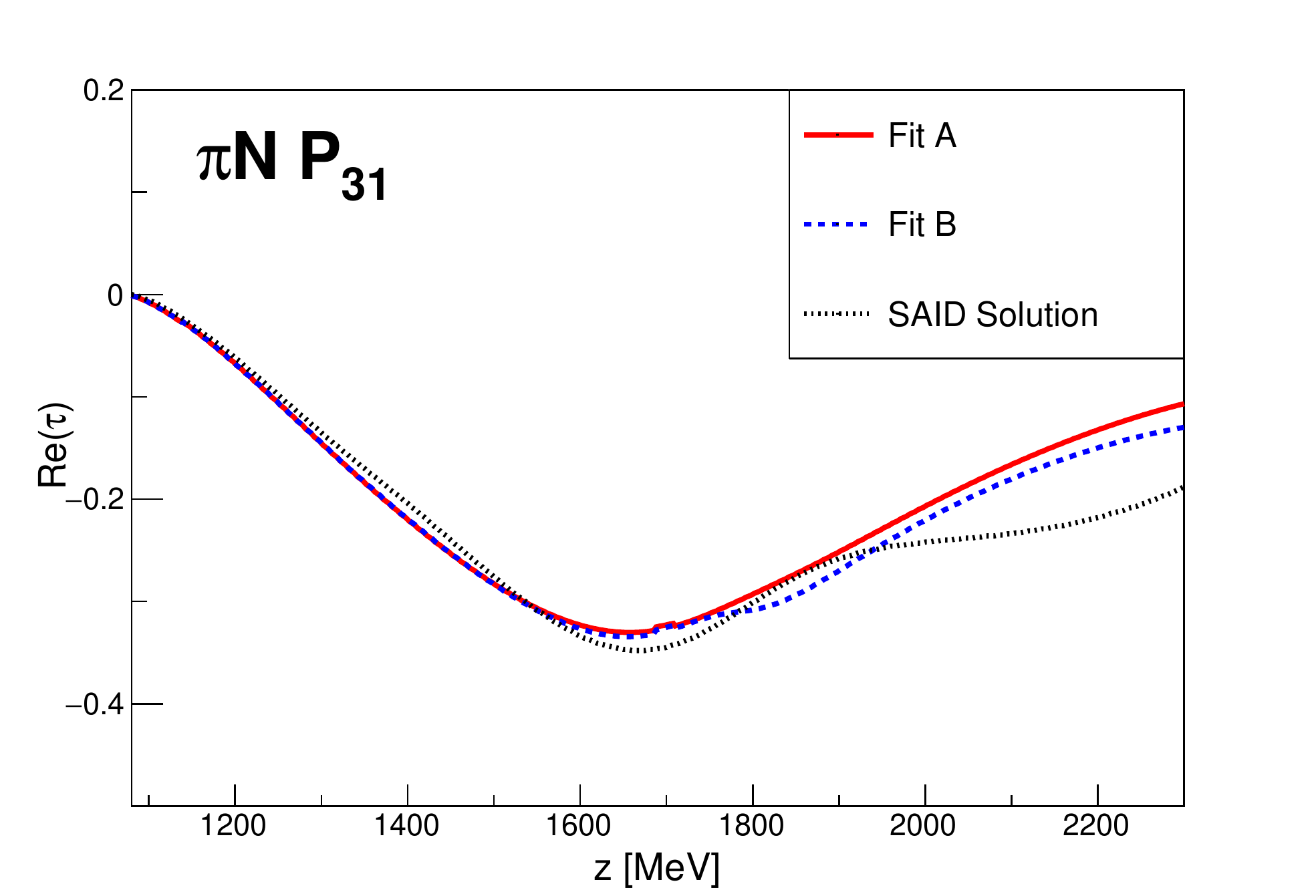}
		\caption{The real part. }
		\label{fig:P31Re}
	\end{subfigure}
	\begin{subfigure}[b]{0.48\textwidth}
		\centering
		\includegraphics[width=1.0\textwidth]{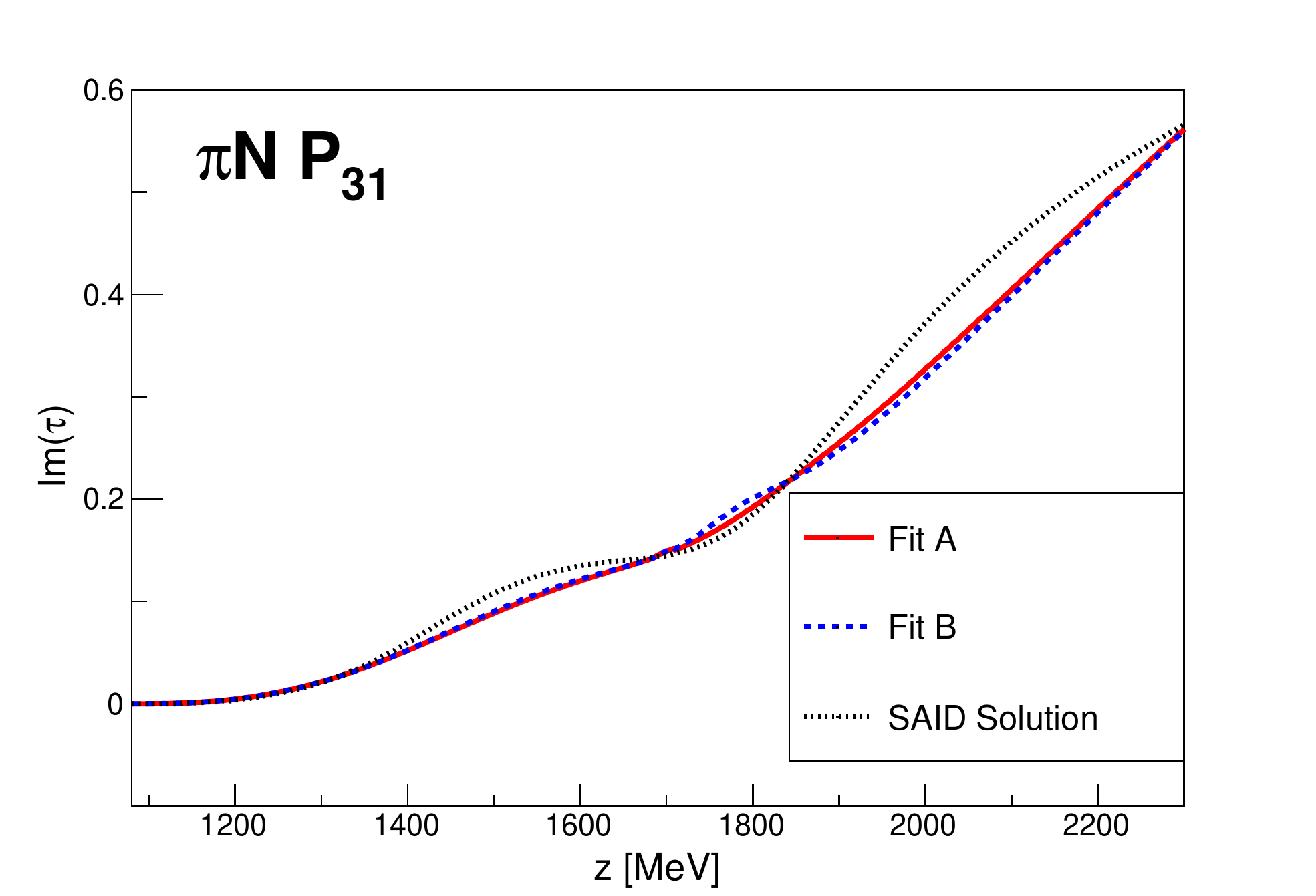}
		\caption{The imaginary part. }
		\label{fig:P31Im}
	\end{subfigure}
	\caption{Fit results of $P_{31}$ $\pi N$ elastic $\tau$ amplitude. The SAID solution is
          from Ref.~\cite{Arndt:2006bf} (energy-dependent solution). We do not need a narrow
          $\Delta(1910)$ pole in both of the two fit scenarios. }\label{fig:P31piN}
\end{figure}

Coming to the $\Delta$ states, even though $\omega N$ is a pure isospin one-half channel, also
the $I=3/2$ parameters are re-fitted as demanded by a coupled-channel framework. Note that
some of the $T^{NP}$ parameters do not only change the isospin $I=1/2$ partial waves and the
amount of the data points for the isospin three-half channels is relatively fewer (only $K^+\Sigma^+$
is purely $I=3/2$). Moreover, the inclusion of the $\omega N$ channel has non-negligible effects
on the $K^+\Sigma^-$ and $K^0\Sigma^0$ final states as the corresponding threshold energies are
close, leading to a rearrangement of the contributions from isospin one-half and three-half.
The states of lower partial waves, e.g. the $\Delta(1620)\,\frac{1}{2}^-$ and $\Delta(1232)\,
\frac{3}{2}^+$, are  more stable. However, the $\Delta(1910)\,\frac{1}{2}^+$, which is of
four-star significance in the PDG~\cite{PDG}, is very broad in this work, just like
Refs.~\cite{Ronchen:2015vfa,Ronchen:2018ury}. It seems that the description of the lineshape in
the $P_{31}$ wave does not need such a significant resonance signal, see Fig.~\ref{fig:P31piN}.
The two states $\Delta(1920)\,\frac{3}{2}^+$ and $\Delta(1940)\,\frac{3}{2}^-$ in our model are
significantly broader than the estimation in PDG~\cite{PDG}. Further, the other states tend to
be narrower. 
\subsubsection{$\omega N$ coupling strengths}
In this section we decompose the normalized residue as $NR=|NR|e^{i\theta}$, with $\theta$ in
units of degrees. Roughly speaking, the modulus $|NR|$ measures how significantly the resonance contributes to the amplitude, while the phase $\theta$ controls the interference behavior among different resonances (and the background). The normalized residues of the $N^*$ states for the $\omega N$ channel are shown
in Tab.~\ref{tab:nroN}. 
\begin{table*}[t!]
	\small
  \begin{ruledtabular}
	\begin{tabular}{cccc} 
		Resonances & Channel (1) & Channel (2)  & Channel (3) \\
		\hline
		$N(1535)\,\frac{1}{2}^-$ & \St{$(1.13,-156^\circ)$\\$(1.13,-163^\circ)$} & $0$ & \St{$(0.14,26^\circ)$\\$(0.13,18^\circ)$}\\ 
    \hline
		$N(1650)\,\frac{1}{2}^-$ & \St{$(0.19,156^\circ)$\\$(0.14,148^\circ)$} & $0$ & \St{$(0.02,-9^\circ)$\\$(0.02,-10^\circ)$} \\
    \hline 
		$N(1440)\,\frac{1}{2}^+$ & \St{$(0.18,-37^\circ)$\\$(0.21,23^\circ)$} & \St{$(0.34,1^\circ)$\\$(0.42,64^\circ)$} & $0$ \\ 
    \hline
		$N(1710)\,\frac{1}{2}^+$ & \St{$(0.10,158^\circ)$\\$(0.27,-86^\circ)$} & \St{$(0.56,-172^\circ)$\\$(0.73,-59^\circ)$} & $0$ \\ 
    \hline
		$N(1880)\,\frac{1}{2}^+$ & \St{$(0.01,-24^\circ)$\\$(0.00,152^\circ)$} & \St{$(0.03,31^\circ)$\\$(0.02,157^\circ)$} & $0$ \\ 
    \hline
		$N(1720)\,\frac{3}{2}^+$ & \St{$(0.01,150^\circ)$\\$(0.01,155^\circ)$} & \St{$(0.05,-178^\circ)$\\$(0.06,-178^\circ)$} & \St{$(0.00,69^\circ)$\\$(0.00,56^\circ)$} \\ 
    \hline
		$N(1900)\,\frac{3}{2}^+$ & \St{$(0.00,33^\circ)$\\$(0.01,-19^\circ)$} & \St{$(0.02,138^\circ)$\\$(0.01,91^\circ)$} & \St{$(0.00,6^\circ)$\\$(0.00,-75^\circ)$} \\ 
    \hline
		$N(1520)\,\frac{3}{2}^-$ & \St{$(0.09,139^\circ)$\\$(0.14,141^\circ)$} & \St{$(0.04,102^\circ)$\\$(0.07,115^\circ)$} & \St{$(0.16,-108^\circ)$\\$(0.22,-99^\circ)$} \\ 
    \hline
		$N(1700)\,\frac{3}{2}^-$ & \St{$(0.02,-35^\circ)$\\$(0.03,20^\circ)$} & \St{$(0.01,-123^\circ)$\\$(0.01,5^\circ)$} & \St{$(0.02,-4^\circ)$\\$(0.01,87^\circ)$} \\ 
    \hline
		$N(1875)\,\frac{3}{2}^-$ & \St{$(0.00,-110^\circ)$\\$(0.00,-82^\circ)$} & \St{$(0.00,-172^\circ)$\\$(0.00,-114^\circ)$} & \St{$(0.00,-157^\circ)$\\$(0.00,-105^\circ)$} \\ 
    \hline
		$N(1675)\,\frac{5}{2}^-$ & \St{$(0.01,108^\circ)$\\$(0.01,117^\circ)$} & \St{$(0.25,82^\circ)$\\$(0.30,89^\circ)$} & \St{$(0.00,-51^\circ)$\\$(0.00,-48^\circ)$} \\  
    \hline
		$N(1680)\,\frac{5}{2}^+$ & \St{$(0.00,-8^\circ)$\\$(0.00,-32^\circ)$} & \St{$(0.04,31^\circ)$\\$(0.04,26^\circ)$} & \St{$(0.95,165^\circ)$\\$(0.98,162^\circ)$} \\ 
    \hline
		$N(1990)\,\frac{7}{2}^+$ & \St{$(0.00,-46^\circ)$\\$(0.00,-42^\circ)$} & \St{$(0.04,-60^\circ)$\\$(0.04,-62^\circ)$} & \St{$(0.00,-105^\circ)$\\$(0.00,-107^\circ)$} \\  
    \hline
		$N(2190)\,\frac{7}{2}^-$ & \St{$(0.00,-155^\circ)$\\$(0.00,-149^\circ)$} & \St{$(0.01,146^\circ)$\\$(0.01,154^\circ)$} & \St{$(0.07,177^\circ)$\\$(0.03,177^\circ)$} \\ 
    \hline
		$N(2250)\,\frac{9}{2}^-$ & \St{$(0.01,-31^\circ)$\\$(0.01,-47^\circ)$} & \St{$(0.12,-28^\circ)$\\$(0.16,-38^\circ)$} & \St{$(0.00,-42^\circ)$\\$(0.01,-52^\circ)$} \\ 
    \hline
		2nd pole~$\frac{9}{2}^-$ & \St{$(0.00,92^\circ)$\\$(0.00,83^\circ)$} & \St{$(0.06,85^\circ)$\\$(0.05,78^\circ)$} & \St{$(0.00,44^\circ)$\\$(0.00,44^\circ)$} \\ 
    \hline
		$N(2220)\,\frac{9}{2}^+$ & \St{$(0.00,50^\circ)$\\$(0.00,58^\circ)$} & \St{$(0.01,10^\circ)$\\$(0.01,14^\circ)$} & \St{$(0.03,21^\circ)$\\$(0.03,24^\circ)$} \\ 
	\end{tabular}
  \end{ruledtabular}
	\caption{The normalized residues of  the $N^*$ states for the $\omega N$ channel. The values
          are written in the form $(NR,\theta)$, with the phase $\theta$ in units of degrees. In
          each cell, the value of the first (second) row is from fit A (B). The three sub-channels
          are: (1)$|J-L|=\frac{1}{2},S=\frac{1}{2}$; (2)$|J-L|=\frac{1}{2},S=\frac{3}{2}$;
          (3)$|J-L|=\frac{3}{2},S=\frac{3}{2}$. }
	\label{tab:nroN}
\end{table*}

In the current fit results, $\omega N$ couples mainly to lower-lying states, and the moduli of the normalized
residues of the $N(1535)\,\frac{1}{2}^-$, $N(1710)\,\frac{1}{2}^+$ and $N(1680)\,\frac{5}{2}^+$ can
be larger than $0.5$. Especially the value for the $N(1535)\,\frac{1}{2}^-$ in channel $(1)$ is
more than one. The large couplings to the $N(1535)\,\frac{1}{2}^-$ and the $N(1710)\,\frac{1}{2}^+$ may stem from the extremely large bare couplings.
At present, we can not determine how model-dependent this result is, and only know that a limited
attempt (the fit~C) has failed: if we force those couplings to be small, then the fit would always
end up in some unphysical local minima with very narrow dynamically generated resonances, and
the description of the data is not satisfactory. Maybe further inclusion of the abundant
photoproduction data can resolve this issue. 

Note that most of the previous studies also support the importance of  the lower states in the
$\omega N$ interaction. For instance, using a quark model, Ref.~\cite{Zhao:2000tb} claims that
the $N(1720)\,\frac{3}{2}^+$ and $N(1680)\,\frac{5}{2}^+$ have the biggest contributions to
$\omega N$ photonproduction\footnote{Unfortunately it is hard to directly compare our result
to Ref.~\cite{Zhao:2000tb}. First, the models are completely different, as we regard hadrons as
the basic degrees of freedom, and second our coupling strengths are actually labelled by normalized
residues.}, while Ref.~\cite{Lutz:2001mi} gets relatively larger couplings of $\omega N$ to
$N(1535)\,\frac{1}{2}^-$, $N(1650)\,\frac{1}{2}^-$ and $N(1520)\,\frac{3}{2}^-$. The coupled-channel
analyses of Refs.~\cite{Penner:2002ma,Penner:2002md,Shklyar:2004ba,Muehlich:2006nn} indicate the
important roles of the $N(1710)\,\frac{1}{2}^+$, $N(1675)\,\frac{5}{2}^-$ and $N(1680)\,\frac{5}{2}^+$.
Extremely huge bare couplings can hardly be avoided in phenomenological coupled-channel models,
e.g. in Ref.~\cite{Muehlich:2006nn}, the large bare couplings of $\omega N$ to $N(1675)\,\frac{5}{2}^-$
and $N(1680)\,\frac{5}{2}^+$ are quite similar to the couplings to $N(1535)\,\frac{1}{2}^-$ and $N(1710)\,\frac{1}{2}^+$ here. Note that we do not associate any physical meaning to the bare couplings.

We also list the nominal branching ratios, defined by Eq.~\eqref{BRdef}, of the states above the
threshold to the $\omega N$ channel, see Tab.~\ref{tab:BRoN}. Though all those resonances do
not couple strongly to $\omega N$, the most significant state is $N(2250)\,\frac{9}{2}^-$, the
branching ratio of which is bigger than $10\%$. Ref.~\cite{Penner:2002md} shows that when 
photoproduction is included, the $N(1900)\,\frac{3}{2}^+$ would play an important role among the
higher resonances, which is different from the hadronic case here. Ref.~\cite{Shklyar:2004ba}
has found the branching ratio of $N(1875)\,\frac{3}{2}^-$\footnote{It is called $D_{13}(1950)$ in
  Ref.~\cite{Shklyar:2004ba}.} to $\omega N$ is large, however, this state is not originally
included in our model and the position of the dynamically generated one is not stable. 
\begin{table*}[t!]
	\small
  \begin{ruledtabular}
	\begin{tabular}{cccc} 
		Resonances & Channel (1) & Channel (2)  & Channel (3)\\
		\hline
		$N(1900)\,\frac{3}{2}^+$ & $(0.09\%)$ & $(0.06\%)$ & $(0.00\%)$\\ 
		$N(1875)\,\frac{3}{2}^-$ & $0.02(0.03)\%$ & $0.00(0.01)\%$ & $0.02(0.01)\%$\\ 
		$N(1990)\,\frac{7}{2}^+$ & $0.01(0.01)\%$ & $3.20(3.16)\%$ & $0.00(0.00)\%$\\  
		$N(2190)\,\frac{7}{2}^-$ & $0.00(0.00)\%$ & $0.09(0.02)\%$ & $3.33(0.79)\%$\\ 
		$N(2250)\,\frac{9}{2}^-$ & $0.02(0.03)\%$ & $12.00(14.23)\%$ & $0.03(0.04)\%$\\  
		2nd pole~$\frac{9}{2}^-$ & $0.02(0.01)\%$ & $6.65(5.82)\%$ & $0.01(0.01)\%$\\ 
		$N(2220)\,\frac{9}{2}^+$ & $0.00(0.00)\%$ & $0.04(0.05)\%$ & $0.80(0.87)\%$\\  
	\end{tabular}
  \end{ruledtabular}
	\caption{The branching ratios of the high-lying $N^*$ states to the $\omega N$ channel
          (in percent). The values outside (inside) the brackets are from fit~A~(B). The $N(1900)\,
          \frac{3}{2}^+$ pole in fit~A is lower than the $\omega N$ threshold. The three sub-channels
          are: (1)$|J-L|=\frac{1}{2},S=\frac{1}{2}$; (2)$|J-L|=\frac{1}{2},S=\frac{3}{2}$;
          (3)$|J-L|=\frac{3}{2},S=\frac{3}{2}$. }
	\label{tab:BRoN}
\end{table*}
\subsubsection{Coupling strengths to the lower channels}
For completeness, we also show the normalized residues of each resonance for the channels
with stable particles lower than $\omega N$. First, the  results for the $N^*$ states are given
in Tab.~\ref{tab:nrlowN}. The pole positions of the $N(1440)\,\frac{1}{2}^+$ and the
$N(1710)\,\frac{1}{2}^+$ in fits~A and B are not so close to each other, so are their residues.
Except for the $N(1535)\,\frac{1}{2}^-$, $N(1650)\,\frac{1}{2}^-$ and $N(1710)\,\frac{1}{2}^+$, all
the other states do not couple strongly to the $KY$ channels. 
\begin{table*}[t!]
	\small
  \begin{ruledtabular}
	\begin{tabular}{ccccc} 
		Resonances & $\pi N$ & $\eta N$ & $K \Lambda$ & $K \Sigma$\\
    \hline
		$N(1535)\,\frac{1}{2}^-$ & \St{$(0.44,-29^\circ)$\\$(0.40,-37^\circ)$} & \St{$(0.48,126^\circ)$\\$(0.48,120^\circ)$} & \St{$(0.65,21^\circ)$\\$(0.66,10^\circ)$} & \St{$(0.21,158^\circ)$\\$(0.18,-171^\circ)$} \\ 
    \hline
		$N(1650)\,\frac{1}{2}^-$ & \St{$(0.46,-58^\circ)$\\$(0.45,-59^\circ)$} & \St{$(0.17,33^\circ)$\\$(0.16,19^\circ)$} & \St{$(0.21,-68^\circ)$\\$(0.21,-74^\circ)$} & \St{$(0.43,-56^\circ)$\\$(0.38,-60^\circ)$} \\ 
    \hline
		$N(1440)\,\frac{1}{2}^+$& \St{$(0.40,-104^\circ)$\\$(0.57,-74^\circ)$} & \St{$(0.03,-62^\circ)$\\$(0.05,26^\circ)$} & \St{$(0.03,102^\circ)$\\$(0.03,178^\circ)$} & \St{$(0.04,-47^\circ)$\\$(0.03,176^\circ)$} \\ 
    \hline
		$N(1710)\,\frac{1}{2}^+$ & \St{$(0.31,-1^\circ)$\\$(0.84,158^\circ)$} & \St{$(0.51,161^\circ)$\\$(0.54,-129^\circ)$} & \St{$(0.41,-23^\circ)$\\$(0.31,104^\circ)$} & \St{$(0.11,79^\circ)$\\$(0.26,-6^\circ)$} \\ 
    \hline
		$N(1880)\,\frac{1}{2}^+$& \St{$(0.01,157^\circ)$\\$(0.02,-43^\circ)$} & \St{$(0.04,-26^\circ)$\\$(0.01,-34^\circ)$} & \St{$(0.04,156^\circ)$\\$(0.01,33^\circ)$} & \St{$(0.07,164^\circ)$\\$(0.09,68^\circ)$} \\ 
    \hline
		$N(1720)\,\frac{3}{2}^+$ & \St{$(0.12,-32^\circ)$\\$(0.13,-28^\circ)$} & \St{$(0.07,120^\circ)$\\$(0.05,102^\circ)$} & \St{$(0.01,-94^\circ)$\\$(0.02,-92^\circ)$} & \St{$(0.01,118^\circ)$\\$(0.03,104^\circ)$} \\ 
    \hline
		$N(1900)\,\frac{3}{2}^+$ & \St{$(0.07,-132^\circ)$\\$(0.15,-163^\circ)$} & \St{$(0.04,53^\circ)$\\$(0.08,28^\circ)$} & \St{$(0.02,-153^\circ)$\\$(0.05,-175^\circ)$} & \St{$(0.00,86^\circ)$\\$(0.00,-75^\circ)$} \\ 
    \hline
		$N(1520)\,\frac{3}{2}^-$ & \St{$(0.67,-11^\circ)$\\$(0.85,-9^\circ)$} & \St{$(0.02,73^\circ)$\\$(0.02,76^\circ)$} & \St{$(0.03,134^\circ)$\\$(0.03,127^\circ)$} & \St{$(0.04,-28^\circ)$\\$(0.01,40^\circ)$} \\ 
    \hline
		$N(1700)\,\frac{3}{2}^-$ & \St{$(0.09,68^\circ)$\\$(0.05,146^\circ)$} & \St{$(0.01,124^\circ)$\\$(0.01,-177^\circ)$} & \St{$(0.01,-148^\circ)$\\$(0.00,-110^\circ)$} & \St{$(0.01,87^\circ)$\\$(0.00,-156^\circ)$} \\ 
    \hline
		$N(1875)\,\frac{3}{2}^-$& \St{$(0.01,-84^\circ)$\\$(0.01,-37^\circ)$} & \St{$(0.00,56^\circ)$\\$(0.00,112^\circ)$} & \St{$(0.00,-171^\circ)$\\$(0.00,-32^\circ)$} & \St{$(0.00,32^\circ)$\\$(0.00,104^\circ)$} \\ 
    \hline
		$N(1675)\,\frac{5}{2}^-$ & \St{$(0.60,-18^\circ)$\\$(0.66,-7^\circ)$} & \St{$(0.13,-56^\circ)$\\$(0.09,-58^\circ)$} & \St{$(0.01,-59^\circ)$\\$(0.02,-9^\circ)$} & \St{$(0.02,-168^\circ)$\\$(0.05,-163^\circ)$} \\ 
    \hline
		$N(1680)\,\frac{5}{2}^+$ & \St{$(0.81,-20^\circ)$\\$(0.79,-21^\circ)$} & \St{$(0.03,83^\circ)$\\$(0.03,91^\circ)$} & \St{$(0.01,-41^\circ)$\\$(0.01,-39^\circ)$} & \St{$(0.01,139^\circ)$\\$(0.00,114^\circ)$} \\ 
    \hline
		$N(1990)\,\frac{7}{2}^+$ & \St{$(0.06,-103^\circ)$\\$(0.06,-104^\circ)$} & \St{$(0.02,125^\circ)$\\$(0.02,140^\circ)$} & \St{$(0.01,-122^\circ)$\\$(0.01,-128^\circ)$} & \St{$(0.01,62^\circ)$\\$(0.01,63^\circ)$} \\ 
    \hline
		$N(2190)\,\frac{7}{2}^-$ & \St{$(0.13,-67^\circ)$\\$(0.14,-67^\circ)$} & \St{$(0.02,92^\circ)$\\$(0.01,-94^\circ)$} & \St{$(0.02,-98^\circ)$\\$(0.01,-112^\circ)$} & \St{$(0.00,-127^\circ)$\\$(0.00,-117^\circ)$} \\ 
    \hline
		$N(2250)\,\frac{9}{2}^-$ & \St{$(0.12,-112^\circ)$\\$(0.19,-128^\circ)$} & \St{$(0.10,-164^\circ)$\\$(0.20,-175^\circ)$} & \St{$(0.02,79^\circ)$\\$(0.01,105^\circ)$} & \St{$(0.03,-161^\circ)$\\$(0.07,-170^\circ)$} \\ 
    \hline
		2nd pole~$\frac{9}{2}^-$ & \St{$(0.06,108^\circ)$\\$(0.04,84^\circ)$} & \St{$(0.06,123^\circ)$\\$(0.03,106^\circ)$} & \St{$(0.00,173^\circ)$\\$(0.00,34^\circ)$} & \St{$(0.01,84^\circ)$\\$(0.01,67^\circ)$} \\ 
    \hline
		$N(2220)\,\frac{9}{2}^+$ & \St{$(0.14,-70^\circ)$\\$(0.14,-69^\circ)$} & \St{$(0.01,112^\circ)$\\$(0.00,-78^\circ)$} & \St{$(0.01,-86^\circ)$\\$(0.01,-89^\circ)$} & \St{$(0.00,-98^\circ)$\\$(0.00,-92^\circ)$} \\ 
	\end{tabular}
  \end{ruledtabular}
	\caption{The normalized residues of the $N^*$ states for the lower channels, given in the
          form $(NR,\theta)$, with the phase $\theta$ in units of degrees. In each cell, the value
          of the first (second) row is from fit~A (B). }
	\label{tab:nrlowN}
\end{table*}

The residues of the $\Delta$ states are summarized in Tab.~\ref{tab:nrlowD}. Apart from those
states that only show up in fit~A or B, the residues of the $\Delta(1910)\,\frac{1}{2}^+$ and
$\Delta(1600)\,\frac{3}{2}^+$ are less stable. As already discussed, the former tends to be
irrelevant to the lineshape in this model, and the latter is affected much by certain $T^{NP}$ parameters. 
\begin{table*}[t!]
	\small
  \begin{ruledtabular}
	\begin{tabular}{ccccc} 
		Resonances & ${\pi N}$ (Fit A) & ${\pi N}$ (Fit B) & ${K \Sigma}$ (Fit A) & ${K \Sigma}$ (Fit B)\\
    \hline
		$\Delta(1620)\,\frac{1}{2}^-$ & $(0.47,-107^\circ)$ & $(0.47,-107^\circ)$ & $(0.19,-104^\circ)$ & $(0.18,-105^\circ)$\\ 
		$\Delta(1750)\,\frac{1}{2}^+$ & $(0.01,144^\circ)$ & $-$ & $(0.03,-81^\circ)$ & $-$\\ 
		$\Delta(1910)\,\frac{1}{2}^+$ & $(0.20,150^\circ)$ & $(0.10,114^\circ)$ & $(0.01,32^\circ)$ & $(0.01,75^\circ)$\\ 
		$\Delta(1232)\,\frac{3}{2}^+$ & $(1.02,-38^\circ)$ & $(1.01,-40^\circ)$ & $(1.12,-169^\circ)$ & $(1.10,-170^\circ)$\\ 
		$\Delta(1600)\,\frac{3}{2}^+$ & $(0.12,-123^\circ)$ & $(0.06,-137^\circ)$ & $(0.14,12^\circ)$ & $(0.07,25^\circ)$\\ 
		$\Delta(1920)\,\frac{3}{2}^+$ & $(0.09,88^\circ)$ & $(0.08,86^\circ)$ & $(0.17,141^\circ)$ & $(0.16,139^\circ)$\\ 
		$\Delta(1700)\,\frac{3}{2}^-$ & $(0.04,-46^\circ)$ & $(0.05,-26^\circ)$ & $(0.01,49^\circ)$ & $(0.01,59^\circ)$\\ 
		$\Delta(1940)\,\frac{3}{2}^-$ & $(0.00,-153^\circ)$ & $(0.00,-157^\circ)$ & $(0.03,22^\circ)$ & $(0.03,19^\circ)$\\ 
		3rd pole~$\frac{3}{2}^-$ & $-$ & $(0.02,141^\circ)$ & $-$ & $(0.01,119^\circ)$\\ 
		$\Delta(1930)\,\frac{5}{2}^-$ & $(0.04,-169^\circ)$ & $(0.05,-153^\circ)$ & $(0.00,-4^\circ)$ & $(0.01,-9^\circ)$\\ 
		$\Delta(1905)\,\frac{5}{2}^+$ & $(0.01,-104^\circ)$ & $(0.04,-98^\circ)$ & $(0.00,-34^\circ)$ & $(0.00,-37^\circ)$\\ 
		$\Delta(1950)\,\frac{7}{2}^+$ & $(0.45,-8^\circ)$ & $(0.45,-8^\circ)$ & $(0.02,-52^\circ)$ & $(0.02,-54^\circ)$\\ 
		$\Delta(2200)\,\frac{7}{2}^-$ & $(0.01,-174^\circ)$ & $(0.09,-160^\circ)$ & $(0.00,-6^\circ)$ & $(0.02,7^\circ)$\\ 
		2nd pole~$\frac{7}{2}^-$ & $-$ & $(0.05,-96^\circ)$ & $-$ & $(0.01,17^\circ)$\\ 
		$\Delta(2400)\,\frac{9}{2}^-$ & $(0.05,-105^\circ)$ & $(0.05,-105^\circ)$ & $(0.00,16^\circ)$ & $(0.00,16^\circ)$\\ 
	\end{tabular}
  \end{ruledtabular}
	\caption{The normalized residues of the $\Delta$ states for the lower channels, written
          in the form $(NR,\theta)$, with the phase $\theta$ in units of degrees. }
	\label{tab:nrlowD}
\end{table*}

The branching ratios to the lower physical channels are also given for the  $N^*$ states in
Tab.~\ref{tab:BRlowN} and for the $\Delta$ states in Tab.~\ref{tab:BRlowD}. We should emphasize
again that the branching ratios here actually come from the residues, and in principle cannot
be directly related to the imaginary part of the pole (the total width). It can happen that
these branching ratios do not sum up to $100\%$, since most of the $N^*$ and $\Delta$ resonances are not typical Breit-Wigner states. 
\begin{table*}[t!]
	\small
  \begin{ruledtabular}
	\begin{tabular}{ccccc} 
		Resonances & $\pi N$ & $\eta N$ & $K\Lambda$ & $K\Sigma$\\
    \hline
		$N(1535)\,\frac{1}{2}^-$ & $44.01(40.04)\%$ & $51.81(56.80)\%$ & $-$ & $-$\\ 
		$N(1650)\,\frac{1}{2}^-$ & $46.93(44.98)\%$ & $6.49(5.89)\%$ & $9.38(9.84)\%$ & $-$\\ 
		$N(1440)\,\frac{1}{2}^+$ & $39.86(56.63)\%$ & $-$ & $-$ & $-$\\ 
		$N(1710)\,\frac{1}{2}^+$ & $30.91(84.47)\%$ & $85.56(34.49)\%$ & $53.29\%$ & $4.17\%$\\ 
		$N(1880)\,\frac{1}{2}^+$ & $1.24(1.89)\%$ & $15.04(0.40)\%$ & $12.53(0.98)\%$ & $38.88(41.85)\%$\\ 
		$N(1720)\,\frac{3}{2}^+$ & $12.15(13.32)\%$ & $3.64(1.55)\%$ & $0.14(0.34)\%$ & $-$\\ 
		$N(1900)\,\frac{3}{2}^+$ & $6.92(15.27)\%$ & $2.09(4.09)\%$ & $0.37(1.62)\%$ & $0.03(0.18)\%$\\ 
		$N(1520)\,\frac{3}{2}^-$ & $66.81(84.65)\%$ & $0.04(0.06)\%$ & $-$ & $-$\\ 
		$N(1700)\,\frac{3}{2}^-$ & $8.72(4.99)\%$ & $-$ & $-$ & $-$\\ 
		$N(1875)\,\frac{3}{2}^-$ & $1.36(1.31)\%$ & $0.02(0.02)\%$ & $0.00(0.00)\%$ & $0.01(0.00)\%$\\ 
		$N(1675)\,\frac{5}{2}^-$ & $59.73(65.74)\%$ & $2.69(1.32)\%$ & $0.03(0.07)\%$ & $-$\\ 
		$N(1680)\,\frac{5}{2}^+$ & $81.17(79.44)\%$ & $0.08(0.08)\%$ & $0.01(0.00)\%$ & $-$\\ 
		$N(1990)\,\frac{7}{2}^+$ & $5.56(5.79)\%$ & $0.58(0.42)\%$ & $0.10(0.18)\%$ & $0.37(0.32)\%$\\ 
		$N(2190)\,\frac{7}{2}^-$ & $12.90(13.69)\%$ & $0.21(0.05)\%$ & $0.17(0.08)\%$ & $0.00(0.00)\%$\\ 
		$N(2250)\,\frac{9}{2}^-$ & $12.20(18.69)\%$ & $7.58(20.42)\%$ & $0.24(0.04)\%$ & $0.97(2.91)\%$\\ 
		2nd pole~$\frac{9}{2}^-$ & $5.58(3.67)\%$ & $5.51(3.08)\%$ & $0.01(0.00)\%$ & $0.35(0.21)\%$\\ 
		$N(2220)\,\frac{9}{2}^+$ & $13.57(13.98)\%$ & $0.03(0.01)\%$ & $0.11(0.07)\%$ & $0.00(0.00)\%$\\ 
	\end{tabular}
  \end{ruledtabular}
	\caption{The branching ratios (in percent) of the $N^*$ states to the lower physical channels,
          defined by Eq.~\eqref{BRdef}. The values outside (inside) the brackets are from fit~A (B). }
	\label{tab:BRlowN}
\end{table*}
\begin{table}[t!]
	\small
  \begin{ruledtabular}
	\begin{tabular}{ccc} 
		Resonances & $\pi N$ & $K\Sigma$\\
		\hline
		$\Delta(1620)\,\frac{1}{2}^-$ & $47.13(46.92)\%$ & $-$ \\ 
		$\Delta(1750)\,\frac{1}{2}^+$ & $1.2\%$ & $5.6\%$ \\ 
		$\Delta(1910)\,\frac{1}{2}^+$ & $19.75(9.93)\%$ & $0.06(0.03)\%$ \\ 
		$\Delta(1232)\,\frac{3}{2}^+$ & $101.86(101.39)\%$ & $-$ \\ 
		$\Delta(1600)\,\frac{3}{2}^+$ & $11.70(6.44)\%$ & $-$ \\ 
		$\Delta(1920)\,\frac{3}{2}^+$ & $9.17(8.04)\%$ & $32.08(32.72)\%$ \\ 
		$\Delta(1700)\,\frac{3}{2}^-$ & $4.46(5.06)\%$ & $0.36(0.33)\%$ \\ 
		$\Delta(1940)\,\frac{3}{2}^-$ & $0.24(0.41)\%$ & $28.86(28.17)\%$ \\ 
		3rd pole~$\frac{3}{2}^-$ & $(2.44\%)$ & $-$ \\ 
		$\Delta(1930)\,\frac{5}{2}^-$ & $3.91(5.20)\%$ & $0.03(0.07)\%$ \\ 
		$\Delta(1905)\,\frac{5}{2}^+$ & $1.33(3.98)\%$ & $0.00(0.00)\%$ \\ 
		$\Delta(1950)\,\frac{7}{2}^+$ & $45.06(45.48)\%$ & $0.12(0.12)\%$ \\ 
		$\Delta(2200)\,\frac{7}{2}^-$ & $0.79(8.75)\%$ & $0.01(0.26)\%$ \\ 
		2nd pole~$\frac{7}{2}^-$ & $(4.80\%)$ & $(0.28\%)$ \\ 
		$\Delta(2400)\,\frac{9}{2}^-$ & $5.17(5.26)\%$ & $0.01(0.01)\%$ \\ 
	\end{tabular}
  \end{ruledtabular}
	\caption{The branching ratios (in percent) of the $\Delta$ states to the lower physical
          channels, defined by Eq.~\eqref{BRdef}. The values outside (inside) the brackets are from
          fit~A (B). }
	\label{tab:BRlowD}
\end{table}

Last, currently for the lack of constraints from $\pi\pi N$ and $\rho N$ data in this model,
the residues for the effective three-body channels are less instructive. Thus, they are
displayed in Appendix~\ref{app:3bodycoup}.  
\subsection{Scattering lengths}
The scattering lengths  are defined as: 
\begin{equation}\label{ascdef}
	a_\kappa\equiv \lim_{p_{\kappa}\to 0} p_{\kappa}^{-1}\tan\tilde{\delta}_\kappa^{(L=0)}\ ,
\end{equation}
where $\kappa$ is the channel label and $\tilde{\delta}_\kappa^{(L=0)}$ is the generalized
$S$-wave phase shift from the diagonal element of the $S$-matrix, 
\begin{equation}\label{deldef}
	S_{\kappa\kappa}\equiv e^{2i\tilde{\delta}_\kappa}\ .
\end{equation}
Note that when the energy is below the $\pi\pi N$ threshold and $\kappa$ corresponds to $\pi N$,
Eq.~\eqref{deldef} is the common definition of the phase shift, namely both $\tilde{\delta}$
and $a$ are real. Specifically the scattering length is extracted from the $\tau$ amplitude
of Eq.~\eqref{taudef}, 
\begin{equation}\label{ascalc}
	a_\kappa=\lim_{p_{\kappa}\to 0}p_{\kappa}^{-1}\tau_{\kappa\kappa}^{(L=0)}\ .
\end{equation}

\begin{figure*}[htpb]
	\centering
	\includegraphics[width=0.7\textwidth]{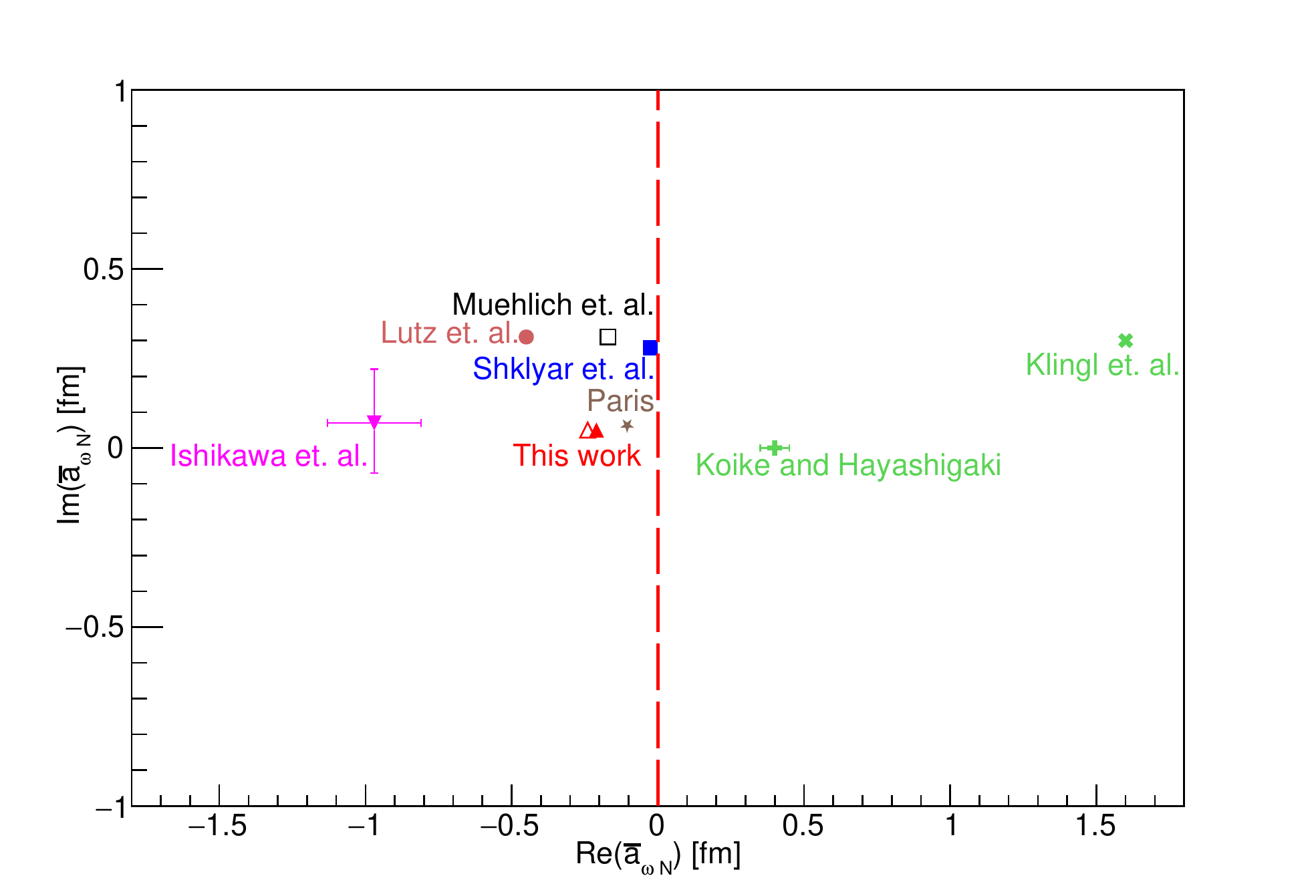}
	\caption{A summary of the spin-averaged $\omega N$ scattering length. The result
          from fit~A~(B) is denoted by the empty (filled) red triangle. The relevant references
          are: Koike and Hayashigaki~\cite{Koike:1996ga}, Klingl et. al.~\cite{Klingl:1998zj},
          Lutz et al.~\cite{Lutz:2001mi}, Shklyar et al.~\cite{Shklyar:2004ba}, Muehlich et
          al.~\cite{Muehlich:2006nn}, Paris~\cite{Paris:2008ig} and Ishikawa et al.~\cite{Ishikawa:2019rvz}. }
	\label{fig:oNasc}
\end{figure*}

As already mentioned, the $\omega N$ scattering length is very important since it indicates
whether the $\omega$ meson can form bound states in the nuclear medium. There are two $\omega N$
scattering lengths  with total spin $S=1/2,3/2$. Here we consider the spin-averaged scattering length,
which is commonly used in the literature (see e.g. Ref.~\cite{Koike:1996ga}): 
\begin{equation}\label{onasc}
	\bar{a}_{\omega N}=\frac{1}{3}a_{\omega N}\Big(S=\frac{1}{2}\Big)+\frac{2}{3}a_{\omega N}\Big(S=\frac{3}{2}\Big)\ .
\end{equation}
The results for fits~A and B are shown in Tab.~\ref{tab:oNasc}. 
\begin{table*}[t!]
	\small
  \begin{ruledtabular}
	\begin{tabular}{cccc} 
		Fit & $a_{\omega N}\Big(S=\frac{1}{2}\Big)$ & $a_{\omega N}\Big(S=\frac{3}{2}\Big)$ & $\bar{a}_{\omega N}$\\
		\hline
		A & $-0.13+0.11i$ & $-0.31+0.01i$ & $-0.24+0.05i$\\ 
		B & $-0.04+0.13i$ & $-0.29+0.01i$ & $-0.21+0.05i$\\ 
	\end{tabular}
  \end{ruledtabular}
	\caption{The scattering lengths of $\omega N$ channel in units of fms. }
	\label{tab:oNasc}
\end{table*}
Though the real parts of the $S=1/2$ scattering length are somewhat different, the two results for
$\bar{a}$ agree with each other qualitatively: both of them have a negative real part, indicating
that our model, irrespective of the uncertainties, does not support bound states of the $\omega$ in
nuclear matter. The imaginary parts are relatively smaller, showing weak inelastic effects near
the $\omega N$ threshold. The values of $\bar{a}_{\omega N}$ here and in the previous studies
are plotted in Fig.~\ref{fig:oNasc}. Our negative real part agrees with all those except for
the two studies based on QCD sum rules. 
Note that in addition to Fig.~\ref{fig:oNasc},
there are some results on the absolute value of the scattering length, e.g. 
$|\bar{a}_{\omega N}|=0.82\pm0.03$~fm in Ref.~\cite{Strakovsky:2014wja}, and also 
$|\bar{a}_{\omega N}|=0.81\pm0.41$~fm from a calculation mentioned therein 
based on Ref.~\cite{Friedman2007}. 

Results for the scattering lengths of the lower channels are shown in Tab.~\ref{tab:otherasc}. The
value of $\pi N$ scattering length is quite similar to Ref.~\cite{Ronchen:2012eg} and compatible
with the result of the Roy-Steiner analyses in Ref.~\cite{Hoferichter:2015hva}, since the inclusion of
a high-lying $\omega N$ channel should not affect the physics of $\pi N$ threshold significantly. However, the scattering lengths of $K\Lambda$ and $K\Sigma\,(I=1/2)$ have changed considerably
compared with the results in Ref.~\cite{Ronchen:2012eg}. This discrepancy possibly stems from
the lack of precision of the corresponding near-threshold data points. The error-bars of the
near-threshold differential cross sections in the $\pi N\to K\Lambda,K\Sigma$ reactions are rather large. 
\begin{table*}[t!]
	\footnotesize
  \begin{ruledtabular}
	\begin{tabular}{ccccc} 
		Result & $a_{\pi N},I=\frac{1}{2}\Big(\frac{3}{2}\Big)$ & $a_{\eta N}$ & $a_{K\Lambda}$
		& $a_{K\Sigma},I=\frac{1}{2}\Big(\frac{3}{2}\Big)$\\
		\hline
		Fit A & $0.25(-0.16)$ & $0.51+0.20i$ & $-0.15+0.05i$ & $-0.01(-0.39)+0.30(0.02)i$\\ 
		Fit B &  $0.25(-0.16)$ & $0.57+0.22i$ & $-0.15+0.04i$ & $-0.07(-0.39)+0.29(0.01)i$\\ 
		Fit A of Ref.~\cite{Ronchen:2012eg} & $0.25(-0.16)$ & $0.49+0.24i$ & $0.04+0.04i$ & $0.36(-0.30)+0.15(0.04)i$\\ 
		Fit B of Ref.~\cite{Ronchen:2012eg} & $0.29(-0.16)$ & $0.55+0.24i$ & $0.04+0.03i$ & $0.32(-0.30)+0.14(0.05)i$\\ 
		Ref.~\cite{Hoferichter:2015hva} & $0.257\pm 0.005(-0.112\pm 0.004)$ & $-$ & $-$ & $-$\\ 
	\end{tabular}
  \end{ruledtabular}
	\caption{The scattering lengths of the lower channels in units of fms. }
	\label{tab:otherasc}
\end{table*}

\section{Conclusion and outlook}\label{sec:con}
In this paper we have performed a refined investigation on pion-induced meson-baryon scattering
reactions, using a sophisticated coupled-channel approach (the J{\"u}lich-Bonn model). The model
includes $t$- and $u$-channel exchange diagrams and $s$-channel genuine states in addition
to phenomenological contact terms, taking into account the $\pi N$, $\pi\pi N$ (effectively
parameterized by $\pi \Delta$, $\sigma N$ and $\rho N$), $\eta N$, $K \Lambda$ and $K \Sigma$
channels with $\omega N$ newly included. This model covers the energy region from the $\pi N$
threshold up to $2.3$~GeV, and fits are done based on all available data. The model-dependence
is estimated by comparing two different fit scenarios. After the calculation of the amplitudes
and extraction of the resonance poles, the $N^*$ and $\Delta$ resonance spectra are reanalysed,
and the following conclusions can be drawn:
\begin{itemize}
\item While the well established $N^*$ states like $N(1535)\,\frac{1}{2}^-$, $N(1650)\,\frac{1}{2}^-$,
  etc., always remain nearly unchanged, some resonances in higher partial waves are found to be stable.
  The main discrepancy of the two fit scenarios is the dynamical structure of the $P_{11}$ wave. 
\item As for the $\Delta$ states, even if $\omega N$ does not couple to them, the results are also
  refined by the new global fits. Specifically, the $\Delta(1910)\,\frac{1}{2}^+$ in our model is
  rather broad. 
\item The $\omega N$ channel mainly couples to three low-lying states in this model, $N(1535)\,
  \frac{1}{2}^-$, $N(1710)\,\frac{1}{2}^+$ and $N(1680)\,\frac{5}{2}^+$. The higher resonances do
  not show markable couplings to the $\omega N$ channel, with the $N(2250)\,\frac{9}{2}^-$ 
  being the most important one. 
\item The spin-averaged scattering length of $\omega N$ is calculated, resulting in
  a negative real part. In fit~A it is $(-0.24+0.05i)\,$fm and in fit~B $(-0.21+0.05i)\,$fm.
  This, in agreement with most other studies, does not support bound states of the $\omega$
  meson in nuclear matter. 
\end{itemize}

There are two main directions for future studies. On the one hand, building on the framework of
Refs.~\cite{Ronchen:2014cna,Ronchen:2022}, we plan to extend the current work to $\omega$ photoproduction,
profiting from the large amount of high-qualtity data that will be of high importance to further
refine the $\omega N$ resonance parameters. On the other hand, the output of this work can be directly
used as the input for e.g. studying the in-medium behavior of the $\omega$ meson, or investigating
the possible hadronic molecules~\cite{Guo:2017jvc} among the $N^*$ and $\Delta$ resonances.
Last, this model can also be employed to the check the possible structures below the $\pi N$
threshold~\cite{Wang:2017agd,Wang:2018gul,Wang:2018nwi,Ma:2020sym,Li:2021tnt,Chen:2022zgm,Cao:2022zhn}. 

\section*{Acknowledgements}
We would like to thank Zhi-Hui Guo, Johann Haidenbauer and Fei Huang for helpful instructions on
consistency checks of the theoretical framework, and Michael D{\"o}ring for useful discussions. 
The authors gratefully acknowledge the computing time 
granted by the JARA Vergabegremium and
provided on the JARA Partition part of the supercomputer JURECA~\cite{JUWELS} at Forschungszentrum
J{\"u}lich. This work is supported by the NSFC and the Deutsche Forschungsgemeinschaft (DFG,
German Research Foundation) through the funds provided to the Sino-German Collaborative Research
Center TRR110 “Symmetries and the Emergence of Structure in QCD” (NSFC Grant No. 12070131001, DFG
Project-ID 196253076-TRR 110) and by the National Natural Science Foundation of China under
Grants No. 12175239 and the National Key R\&D Program of China under Contract No. 2020YFA0406400.
Further support by the CAS through a President’s International Fellowship Initiative (PIFI)
(Grant No. 2018DM0034) and by the VolkswagenStiftung (Grant No. 93562) is acknowledged.

\appendix
\section{Observables of the omega-nucleon channel}\label{app:expr}
The expression of the differential cross section for either $\pi^-p\to\omega n$ or
$\pi^+n\to\omega p$ is: 
\begin{widetext}
\begin{equation}\label{dsigomgN}
\begin{split}
	&\frac{d\bar{\sigma}}{d\Omega}
	=\frac{2}{3p_i^2}\Bigg[
	\Big|\sum_J(2J+1)\tau^J_{++0}d^J_{1/2,1/2}(\theta)\Big|^2
	+\Big|\sum_J(2J+1)\tau^J_{+-0}d^J_{1/2,-1/2}(\theta)\Big|^2\\
	&+\Big|\sum_J(2J+1)\tau^J_{+--}d^J_{1/2,1/2}(\theta)\Big|^2
	+\Big|\sum_J(2J+1)\tau^J_{+++}d^J_{1/2,-1/2}(\theta)\Big|^2\\
	&+\Big|\sum_J(2J+1)\tau^J_{++-}d^J_{1/2,3/2}(\theta)\Big|^2
	+\Big|\sum_J(2J+1)\tau^J_{+-+}d^J_{1/2,-3/2}(\theta)\Big|^2\Bigg]\ , 
\end{split}
\end{equation}
\end{widetext}
where the initial spins are averaged and the final ones are summed, the $d^J$'s are Wigner $d$-functions (with the scattering angle $\theta$ as the argument) and $\tau^J_{\lambda_1\lambda_3\lambda_4}$
are the amplitudes of Eq.~\eqref{taudef} written in the helicity basis (total isospin $I=1/2$),
with $\lambda_1,\lambda_3$ and $\lambda_4$ being the helicities of initial nucleon, final nucleon
and final $\omega$, respectively (``$\pm$'' stands for $\lambda=\pm \frac{1}{2}$ or $\pm 1$).
The relation between the $\tau$ amplitudes in Eq.~\eqref{taudef} ($JLS$ basis) and in
Eq.~\eqref{dsigomgN} (helicity basis) is: 
\begin{widetext}
\begin{equation}
    \begin{pmatrix}
        \tau^J_{++0} \\ \tau^J_{+-0} \\ \tau^J_{+--} \\ \tau^J_{+++} \\ \tau^J_{++-} \\ \tau^J_{+-+}
    \end{pmatrix}
    =\frac{1}{12}
    \begin{pmatrix}
        2\sqrt{3} & -3\sqrt{\frac{2J-1}{J}} & \sqrt{\frac{6J+9}{J}} & 2\sqrt{3} & 3\sqrt{\frac{2J+3}{J+1}} & -\sqrt{\frac{6J-3}{J+1}}\\
        2\sqrt{3} & -3\sqrt{\frac{2J-1}{J}} & \sqrt{\frac{6J+9}{J}} & -2\sqrt{3} & -3\sqrt{\frac{2J+3}{J+1}} & \sqrt{\frac{6J-3}{J+1}}\\
        -2\sqrt{6} & -3\sqrt{\frac{2J-1}{2J}} & \sqrt{\frac{6J+9}{2J}} & -2\sqrt{6} & 3\sqrt{\frac{2J+3}{2J+2}} & -\sqrt{\frac{6J-3}{2J+2}}\\
        -2\sqrt{6} & -3\sqrt{\frac{2J-1}{2J}} & \sqrt{\frac{6J+9}{2J}} & 2\sqrt{6} & -3\sqrt{\frac{2J+3}{2J+2}} & \sqrt{\frac{6J-3}{2J+2}}\\
        0 & -3\sqrt{\frac{2J+3}{2J}} & -3\sqrt{\frac{6J-3}{2J}} & 0 & -3\sqrt{\frac{2J-1}{2J+2}} & -3\sqrt{\frac{6J+9}{2J+2}}\\
        0 & -3\sqrt{\frac{2J+3}{2J}} & -3\sqrt{\frac{6J-3}{2J}} & 0 & 3\sqrt{\frac{2J-1}{2J+2}} & 3\sqrt{\frac{6J+9}{2J+2}}
    \end{pmatrix}
    \begin{pmatrix}
        \tau^J_{p_1p_1S_1} \\ \tau^J_{p_1m_3S_3} \\ \tau^J_{p_1p_1S_3} \\ \tau^J_{m_1m_1S_1} \\ \tau^J_{m_1p_3S_3} \\ \tau^J_{m_1m_1S_3}
    \end{pmatrix}\ ,
\end{equation}
\end{widetext}
using the abbreviations
\begin{equation}
    \begin{pmatrix}
        \tau^J_{p_1p_1S_1} \\ \tau^J_{p_1m_3S_3} \\ \tau^J_{p_1p_1S_3} \\ \tau^J_{m_1m_1S_1} \\ \tau^J_{m_1p_3S_3} \\ \tau^J_{m_1m_1S_3}
    \end{pmatrix}
    \equiv
    \begin{pmatrix}
        \tau^J\Big(L_i=J+\frac{1}{2},L_f=J+\frac{1}{2},S_f=\frac{1}{2}\Big)\\ 
        \tau^J\Big(L_i=J+\frac{1}{2},L_f=J-\frac{3}{2},S_f=\frac{3}{2}\Big)\\ 
        \tau^J\Big(L_i=J+\frac{1}{2},L_f=J+\frac{1}{2},S_f=\frac{3}{2}\Big)\\ 
        \tau^J\Big(L_i=J-\frac{1}{2},L_f=J-\frac{1}{2},S_f=\frac{1}{2}\Big)\\ 
        \tau^J\Big(L_i=J-\frac{1}{2},L_f=J+\frac{3}{2},S_f=\frac{3}{2}\Big)\\ 
        \tau^J\Big(L_i=J-\frac{1}{2},L_f=J-\frac{1}{2},S_f=\frac{3}{2}\Big)
    \end{pmatrix}\ .
\end{equation}

After integrating over the angular dependence, the total cross section for
either $\pi^-p\to\omega n$ or $\pi^+n\to\omega p$ is: 
\begin{equation}\label{totXsec}
\begin{split}
    \bar{\sigma}&=\frac{8\pi}{3p_i^2}\sum_{J} (2J+1)\Big(|\tau^J_{++0}|^2+|\tau^J_{+-0}|^2+|\tau^J_{+--}|^2\\
    &+|\tau^J_{+++}|^2+|\tau^J_{++-}|^2+|\tau^J_{+-+}|^2\Big)\ .
\end{split}
\end{equation}

\section{Partial wave amplitudes of $\pi N\to\omega N$}\label{app:oNpw}
\begin{figure}[t!]
	\centering
	\includegraphics[width=0.45\textwidth]{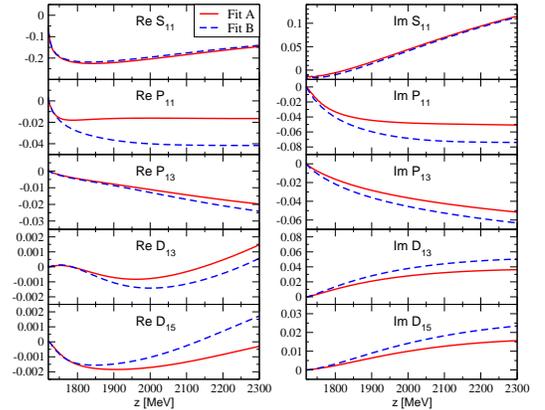}
	\caption{Partial wave amplitudes with $L_{\pi N}\leq 2$, in the sub-channel~(1) of $\omega N$: $|J-L_{\omega N}|=\frac{1}{2},S_{\omega N}=\frac{1}{2}$. The notation $L_{2I,2J}$ is for the
          initial $\pi N$ system. }
	\label{fig:oNpwlow1}
\end{figure}

\begin{figure}[t!]
	\centering
	\includegraphics[width=0.45\textwidth]{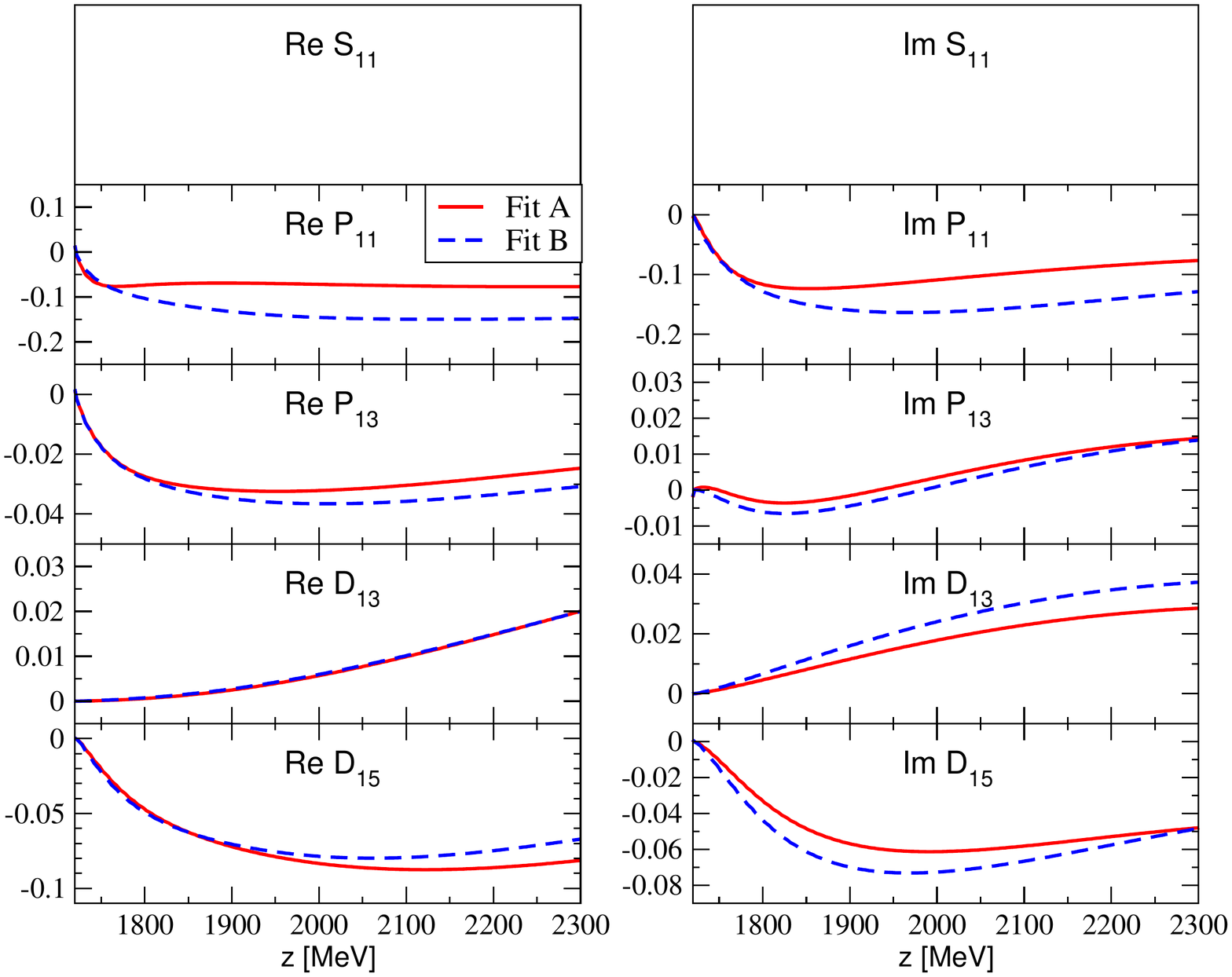}
	\caption{Partial wave amplitudes with $L_{\pi N}\leq 2$, in the sub-channel~(2) of $\omega N$: $|J-L_{\omega N}|=\frac{1}{2},S_{\omega N}=\frac{3}{2}$. The notation $L_{2I,2J}$ is for the
          initial $\pi N$ system. }
	\label{fig:oNpwlow2}
\end{figure}

\begin{figure}[t!]
	\centering
	\includegraphics[width=0.45\textwidth]{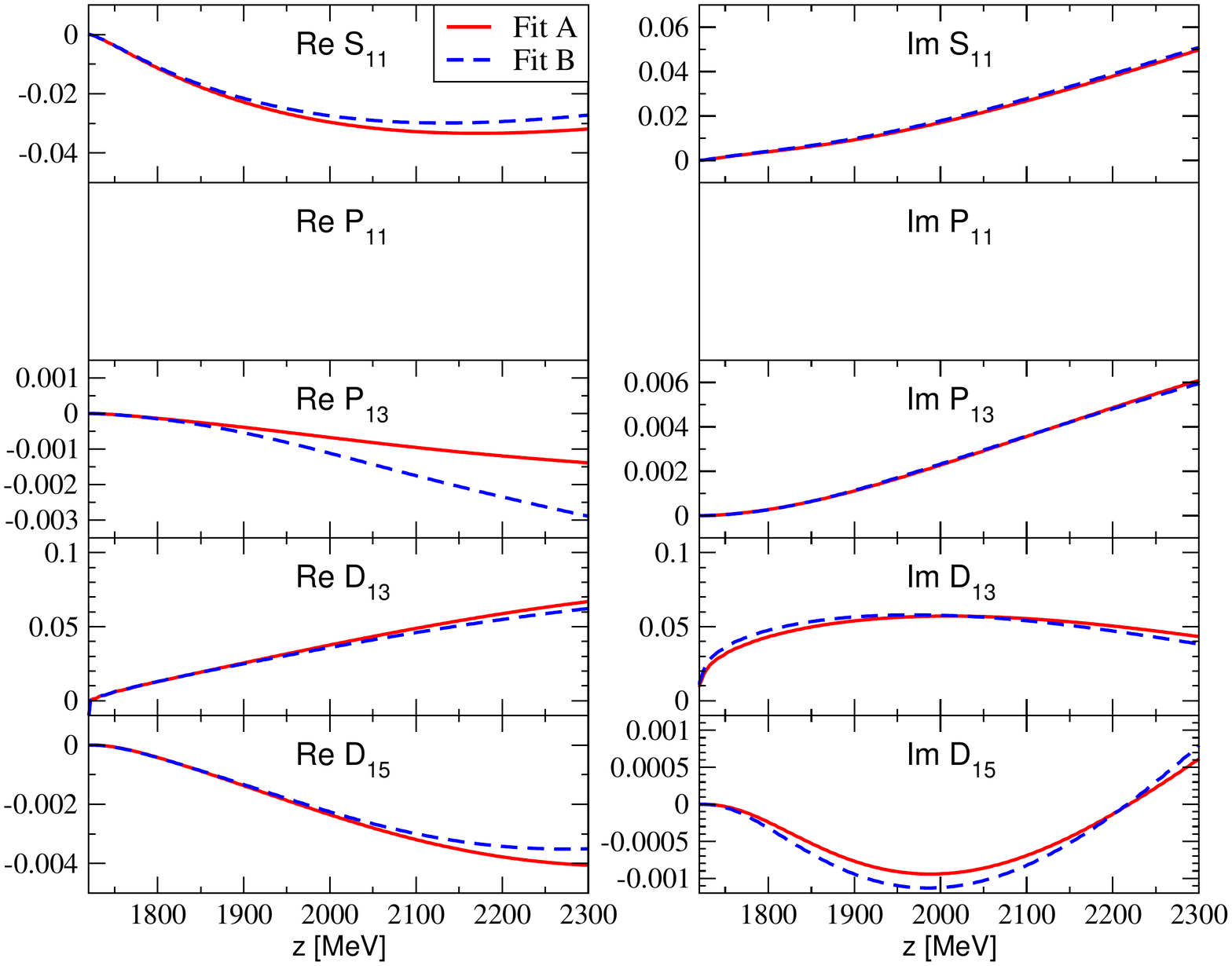}
	\caption{Partial wave amplitudes with $L_{\pi N}\leq 2$, in the sub-channel~(3) of $\omega N$: $|J-L_{\omega N}|=\frac{3}{2},S_{\omega N}=\frac{3}{2}$. The notation $L_{2I,2J}$ is for the
          initial $\pi N$ system. }
	\label{fig:oNpwlow3}
\end{figure}

\begin{figure}[t!]
	\centering
	\includegraphics[width=0.45\textwidth]{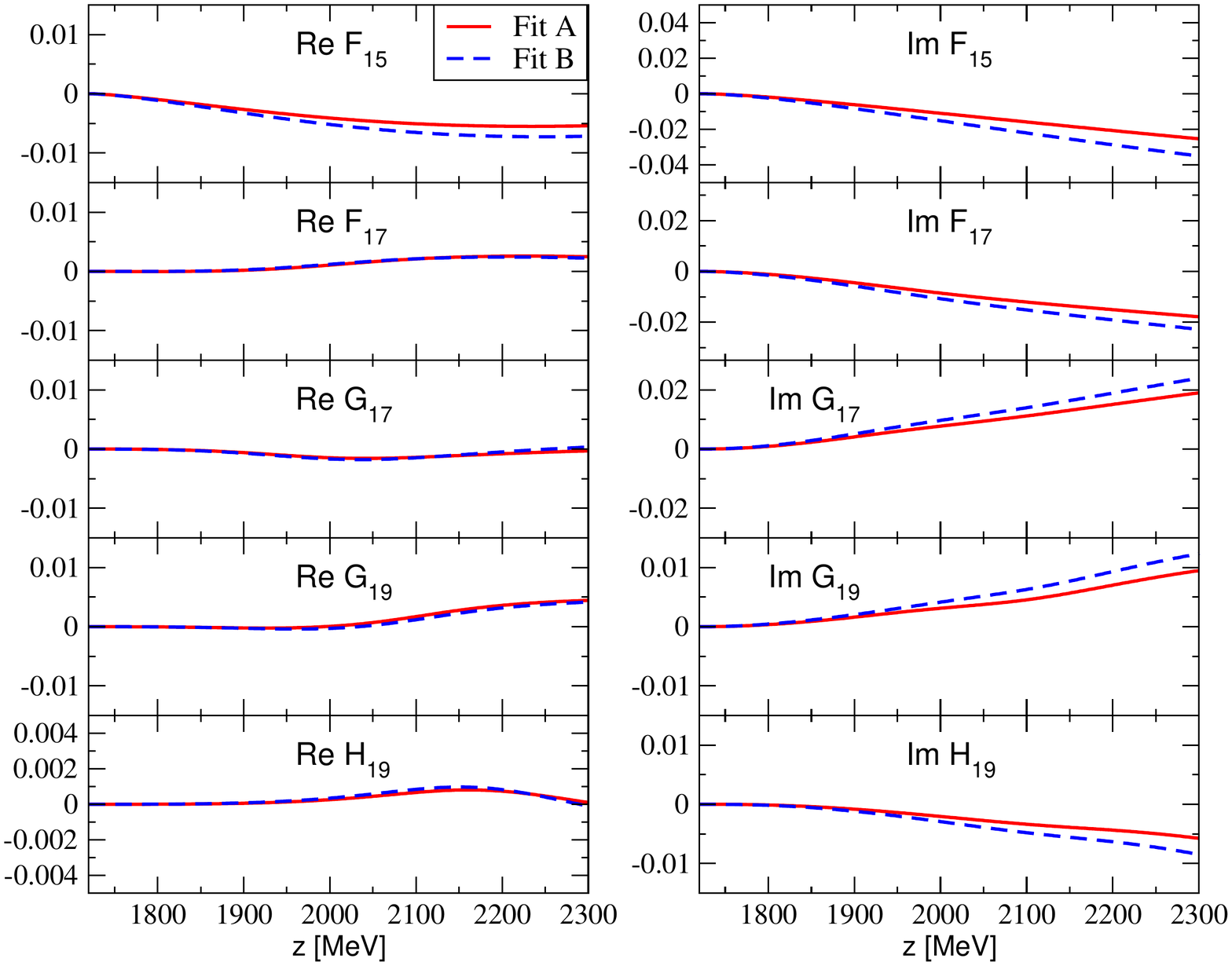}
	\caption{Partial wave amplitudes with $L_{\pi N}\geq 3$, in the sub-channel~(1) of $\omega N$:
          $|J-L_{\omega N}|=\frac{1}{2},S_{\omega N}=\frac{1}{2}$. The notation $L_{2I,2J}$ is for the
          initial $\pi N$ system. }
	\label{fig:oNpwhigh1}
\end{figure}

\begin{figure}[t!]
	\centering
	\includegraphics[width=0.45\textwidth]{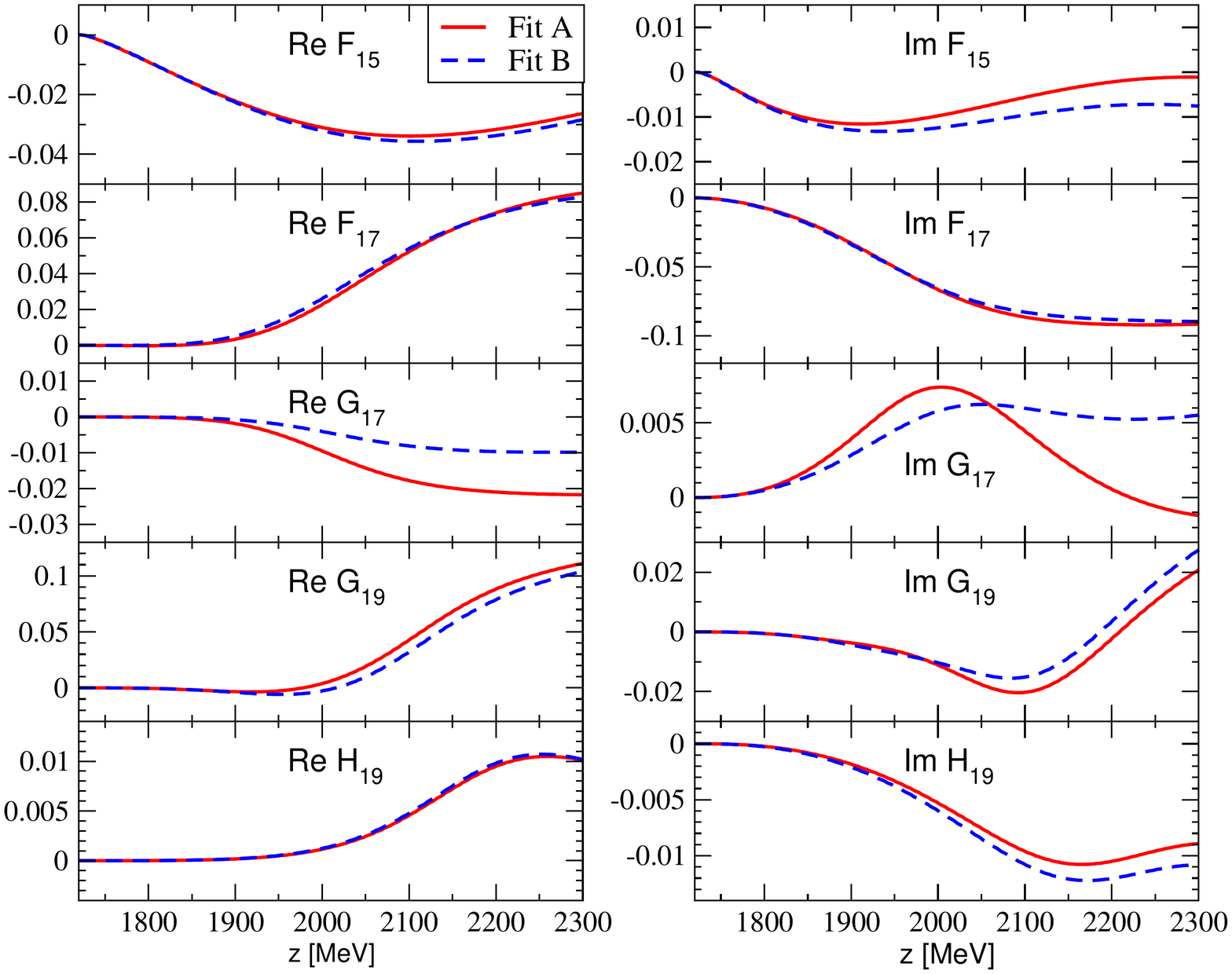}
	\caption{Partial wave amplitudes with $L_{\pi N}\geq 3$, in the sub-channel~(2) of $\omega N$:
          $|J-L_{\omega N}|=\frac{1}{2},S_{\omega N}=\frac{3}{2}$. The notation $L_{2I,2J}$ is for the
          initial $\pi N$ system. }
	\label{fig:oNpwhigh2}
\end{figure}

\begin{figure}[t!]
	\centering
	\includegraphics[width=0.45\textwidth]{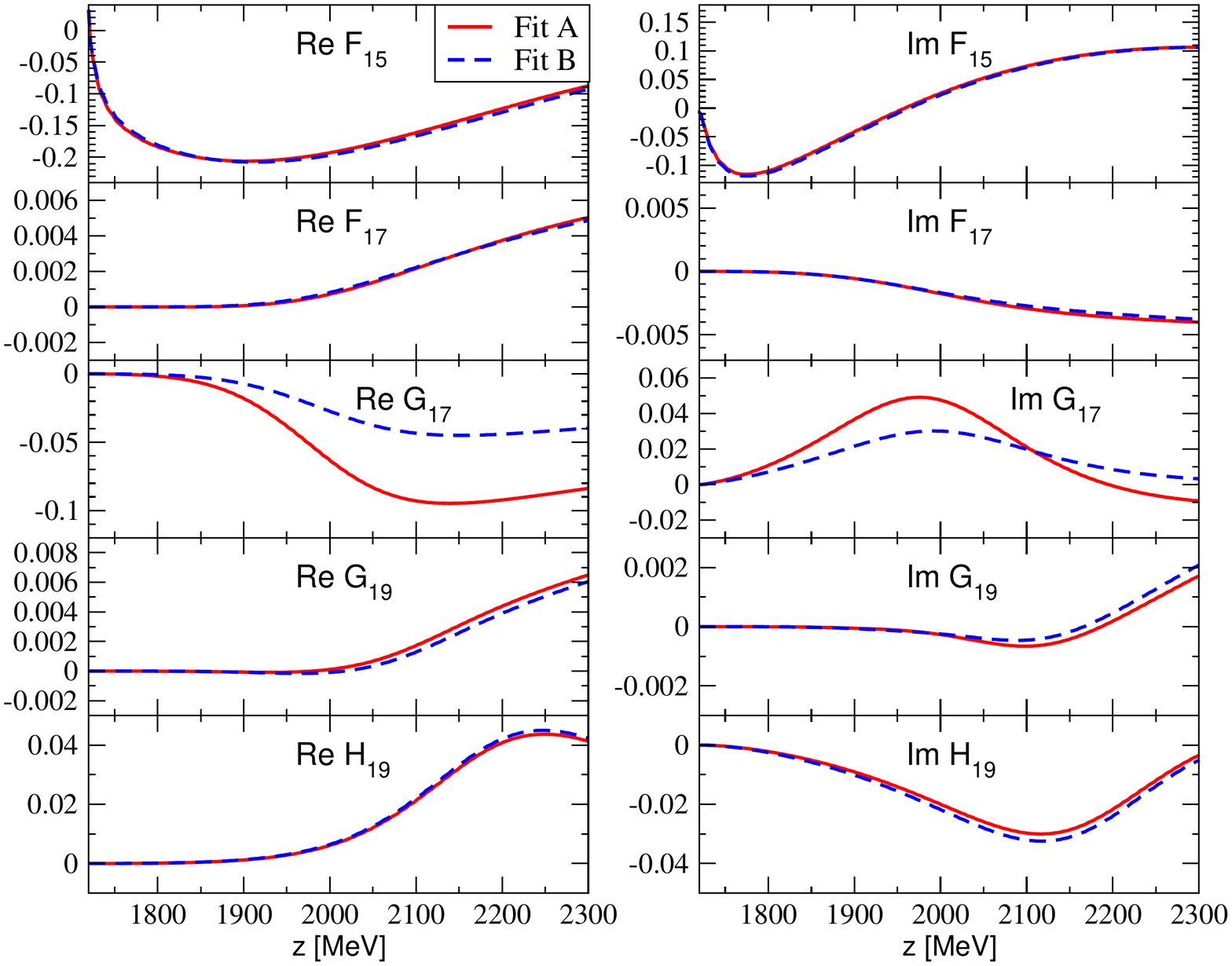}
	\caption{Partial wave amplitudes with $L_{\pi N}\geq 3$, in the sub-channel~(3) of $\omega N$:
          $|J-L_{\omega N}|=\frac{3}{2},S_{\omega N}=\frac{3}{2}$. The notation $L_{2I,2J}$ is for the
          initial $\pi N$ system. }
	\label{fig:oNpwhigh3}
\end{figure}

The partial wave amplitudes of $\pi N\to\omega N$ in this work do not show significant structures, and depend not much on the two different fit solutions especially when the energy is close to the $\omega N$ threshold.
For partial waves with ($\pi N$) orbital angular momentum $L\leq 2$, see
Figs.~\ref{fig:oNpwlow1}, \ref{fig:oNpwlow2} and \ref{fig:oNpwlow3}, for $L\geq 3$
see Figs.~\ref{fig:oNpwhigh1}, \ref{fig:oNpwhigh2} and \ref{fig:oNpwhigh3}. The
magnitude of $F_{15}$ amplitude in sub-channel~(3) is relatively large, since the coupling of
$\omega N$ to $N(1680)\,\frac{5}{2}^+$ is rather strong. Note that the $N^*(1680)$ is only
a $P$-wave resonance for $\omega N$. 
\section{Couplings of the resonances to effective three-body channels}\label{app:3bodycoup}
In this section the complex couplings $g_\mu$ to the effective three-body channels
($\sigma N$ excluded) are given as: 
\begin{equation}\label{Tpole}
	\tilde{T}_{\mu\nu}\sim \frac{g_\mu g_\nu}{z-z_0}+\cdots\ ;
\end{equation}
where $g_\mu$ is related to $R_\mu$ by Eq.~\eqref{taudef}. The results are shown for the $N^*$
states in Tab.~\ref{tab:threebodyNst} and for the $\Delta$ states in Tab.~\ref{tab:threebodyDel}.
In this model we do not consider the couplings of $\sigma N$ to the resonances. 
\begin{table*}[b!]
	\scriptsize
	\begin{ruledtabular}
	\begin{tabular}{cccccc} 
		Resonances & $\pi \Delta$ (1) & $\pi \Delta$ (2) & $\rho N$ (1) & $\rho N$ (2) & $\rho N$ (3)\\
		\hline
		$N(1535)\,\frac{1}{2}^-$ & $0$ & \St{$-6.3+4.2i$\\$(-6.0+4.4i)$} & \St{$-5.4-1.2i$\\$(-4.9-0.8i)$} & $0$ & \St{$-0.2-0.4i$\\$(-0.2-0.3i)$}\\ 
		\hline
		$N(1650)\,\frac{1}{2}^-$ & $0$ & \St{$4.1+4.1i$\\$(5.1+3.3i)$} & \St{$-2.2+16.1i$\\$(-1.5+15.5i)$} & $0$ & \St{$0.4+2.6i$\\$(0.5+2.6i)$}\\ 
		\hline
		$N(1440)\,\frac{1}{2}^+$ & \St{$-6.7+4.3i$\\$(-7.2+4.4i)$} & $0$ & \St{$1.1+3.1i$\\$(-1.4+5.6i)$} & \St{$3.6-3.2i$\\$(4.3-0.7i)$} & $0$\\ 
		\hline
		$N(1710)\,\frac{1}{2}^+$ & \St{$8.6+3.7i$\\$(11.6-5.8i)$} & $0$ & \St{$-1.4+0.1i$\\$(3.5+5.0i)$} & \St{$1.5-1.0i$\\$(4.9-4.4i)$} & $0$\\
		\hline 
		$N(1880)\,\frac{1}{2}^+$ & \St{$2.8+0.2i$\\$(1.1+0.5i)$} & $0$ & \St{$0.1+1.0i$\\$(-0.7+0.6i)$} & \St{$0.2-0.2i$\\$(0.5+0.0i)$} & $0$\\ 
		\hline
		$N(1720)\,\frac{3}{2}^+$ & \St{$14.4+2.7i$\\$(16.5+3.3i)$} & \St{$-0.6-0.1i$\\$(-0.6-0.2i)$} & \St{$-1.3-1.0i$\\$(-1.5-1.2i)$} & \St{$-2.5+13.8i$\\$(-2.3+14.9i)$} & \St{$1.0-0.4i$\\$(1.1-0.5i)$}\\ 
		\hline
		$N(1900)\,\frac{3}{2}^+$ & \St{$-9.1+11.3i$\\$(-7.7+20.7i)$} & \St{$-0.1-0.2i$\\$(-2.7+1.2i)$} & \St{$1.6-0.4i$\\$(2.3-0.7i)$} & \St{$2.6+4.2i$\\$(5.8+5.0i)$} & \St{$0.3-1.8i$\\$(0.4-1.1i)$}\\ 
		\hline
		$N(1520)\,\frac{3}{2}^-$ & \St{$-0.1+1.4i$\\$(-0.1+1.7i)$} & \St{$-13.2-2.5i$\\$(-14.0-4.0i)$} & \St{$-1.8+1.2i$\\$(-2.1+1.2i)$} & \St{$-1.7+0.2i$\\$(-1.9+0.2i)$} & \St{$2.4-24.6i$\\$(3.0-28.0i)$}\\ 
		\hline
		$N(1700)\,\frac{3}{2}^-$ & \St{$0.3+1.6i$\\$(1.0-1.0i)$} & \St{$16.0-15.1i$\\$(-21.8+14.2i)$} & \St{$-2.4-1.7i$\\$(3.7+2.5i)$} & \St{$0.9+0.5i$\\$(-1.1+3.9i)$} & \St{$-1.7+0.0i$\\$(-3.1-8.9i)$}\\ 
		\hline
		$N(1875)\,\frac{3}{2}^-$ & \St{$-12.1+10.1i$\\$(-13.5+7.1i)$} & \St{$1.9-0.2i$\\$(2.7+0.7i)$} & \St{$-2.4-0.2i$\\$(-2.1-1.1i)$} & \St{$0.9-1.2i$\\$(1.1-1.2i)$} & \St{$0.5+0.9i$\\$(0.3+0.6i)$}\\ 
		\hline
		$N(1675)\,\frac{5}{2}^-$ & \St{$-12.9+3.2i$\\$(-12.5+1.8i)$} & \St{$0.3-0.0i$\\$(0.3+0.0i)$} & \St{$1.5+2.3i$\\$(1.0+2.2i)$} & \St{$-4.3+0.6i$\\$(-4.4-0.1i)$} & \St{$-1.0-0.2i$\\$(-1.1-0.2i)$}\\ 
		\hline
		$N(1680)\,\frac{5}{2}^+$ & \St{$-0.2-0.1i$\\$(-0.2-0.0i)$} & \St{$-6.3+1.4i$\\$(-6.5+1.2i)$} & \St{$-1.2+1.0i$\\$(-1.2+1.0i)$} & \St{$-0.7+0.5i$\\$(-0.7+0.4i)$} & \St{$0.0-4.4i$\\$(0.1-4.4i)$}\\ 
		\hline
		$N(1990)\,\frac{7}{2}^+$ & \St{$-5.0+5.0i$\\$(-5.3+5.1i)$} & \St{$-0.0-0.9i$\\$(-0.3-1.2i)$} & \St{$1.1+0.6i$\\$(1.2+0.6i)$} & \St{$6.1+4.0i$\\$(6.3+4.1i)$} & \St{$0.1-0.0i$\\$(0.1-0.1i)$}\\ 
		\hline
		$N(2190)\,\frac{7}{2}^-$ & \St{$0.3-1.0i$\\$(0.3-1.0i)$} & \St{$-6.7+6.6i$\\$(-6.7+6.9i)$} & \St{$-0.2-0.1i$\\$(-0.3-0.1i)$} & \St{$-0.4-0.9i$\\$(-0.5-1.0i)$} & \St{$-3.1-4.9i$\\$(-3.2-5.3i)$}\\ 
		\hline
		$N(2250)\,\frac{9}{2}^-$ & \St{$-7.1+5.4i$\\$(-8.5+6.5i)$} & \St{$0.7-0.5i$\\$(0.7-0.6i)$} & \St{$1.1+0.1i$\\$(1.0+0.3i)$} & \St{$-6.2-3.0i$\\$(-7.5-3.9i)$} & \St{$-0.9+0.3i$\\$(-0.9+0.5i)$}\\ 
		\hline
		2nd pole~$\frac{9}{2}^-$ & \St{$-7.3-0.1i$\\$(-6.7+0.8i)$} & \St{$0.5-0.1i$\\$(0.4-0.0i)$} & \St{$-0.5+0.5i$\\$(0.3+0.7i)$} & \St{$-4.9-0.8i$\\$(-4.2-1.4i)$} & \St{$0.3+0.3i$\\$(-0.4-0.1i)$}\\ 
		\hline
		$N(2220)\,\frac{9}{2}^+$ & \St{$-0.2+0.4i$\\$(-0.3+0.4i)$} & \St{$10.3-5.3i$\\$(10.4-5.2i)$} & \St{$-0.9-1.5i$\\$(-1.0-1.5i)$} & \St{$-0.5-1.8i$\\$(-0.5-1.8i)$} & \St{$2.6+0.3i$\\$(2.6+0.3i)$}\\ 
	\end{tabular}
	\end{ruledtabular}
	\caption{The complex couplings of $N^*$ states to the effective three-body channels
          (in units of $10^{-3}\sqrt{\text{MeV}}$), defined by Eq.~\eqref{Tpole}. The values
          outside (inside) the brackets are from fit A (B). $\pi\Delta$ (1,2) stand for $|J-L|=\frac{1}{2},\frac{3}{2}$, respectively; while the three sub-channels of $\rho N$ are: (1)$|J-L|=\frac{1}{2},S=\frac{1}{2}$; (2)$|J-L|=\frac{1}{2},S=\frac{3}{2}$; (3)$|J-L|=\frac{3}{2},S=\frac{3}{2}$. }
	\label{tab:threebodyNst}
\end{table*}
\begin{table*}[b!]
	\scriptsize
	\begin{ruledtabular}
	\begin{tabular}{cccccc} 
		Resonances & $\pi \Delta$ (1) & $\pi \Delta$ (2) & $\rho N$ (1) & $\rho N$ (2) & $\rho N$ (3)\\
		\hline
		$\Delta(1620)\,\frac{1}{2}^-$ & $0$ & \St{$-10.1+3.4i$\\$(-10.1+3.3i)$} & \St{$2.5+7.6i$\\$(2.4+7.2i)$} & $0$ & \St{$0.7+0.9i$\\$(0.7+0.9i)$}\\ 
		\hline
		$\Delta(1750)\,\frac{1}{2}^+$ & $-1.2-0.7i$ & $0$ & $0.5+1.9i$ & $0.6+0.6i$ & $0$\\ 
		\hline
		$\Delta(1910)\,\frac{1}{2}^+$ & \St{$-15.3+12.8i$\\$(5.6-11.1i)$} & $0$ & \St{$-1.7-1.2i$\\$(1.4+0.0i)$} & \St{$-1.4-4.1i$\\$(3.3+0.8i)$} & $0$\\ 
		\hline
		$\Delta(1232)\,\frac{3}{2}^+$ & \St{$54.8+20.0i$\\$(59.5+23.9i)$} & \St{$-4.3+2.9i$\\$(-5.8+2.2i)$} & \St{$-8.1-9.3i$\\$(-7.9-9.0i)$} & \St{$1.3-2.7i$\\$(0.4-4.3i)$} & \St{$6.7+2.2i$\\$(6.4+1.8i)$}\\ 
		\hline
		$\Delta(1600)\,\frac{3}{2}^+$ & \St{$-16.4+6.6i$\\$(-10.4+5.8i)$} & \St{$0.5-0.0i$\\$(0.3-0.0i)$} & \St{$7.5+2.4i$\\$(4.8+0.7i)$} & \St{$7.4+5.4i$\\$(4.7+3.0i)$} & \St{$-2.4+0.2i$\\$(-1.5+0.8i)$}\\ 
		\hline
		$\Delta(1920)\,\frac{3}{2}^+$ & \St{$3.0-0.4i$\\$(3.5-0.0i)$} & \St{$-1.1-0.7i$\\$(-1.1-0.6i)$} & \St{$0.2+2.3i$\\$(0.4+2.0i)$} & \St{$1.0+0.9i$\\$(1.2+0.9i)$} & \St{$-1.5-0.1i$\\$(-1.4-0.1i)$}\\ 
		\hline
		$\Delta(1700)\,\frac{3}{2}^-$ & \St{$1.9+2.9i$\\$(2.0+2.6i)$} & \St{$-19.1+4.3i$\\$(-20.1+4.2i)$} & \St{$0.3-1.7i$\\$(0.2-2.1i)$} & \St{$2.2-0.2i$\\$(2.3-0.4i)$} & \St{$7.2+7.6i$\\$(7.6+8.1i)$}\\ 
		\hline
		$\Delta(1940)\,\frac{3}{2}^-$ & \St{$-2.1+0.5i$\\$(-1.9+0.4i)$} & \St{$0.8-2.8i$\\$(0.7-2.0i)$} & \St{$-0.3+0.4i$\\$(-0.3+0.3i)$} & \St{$-0.0+0.3i$\\$(-0.0+0.2i)$} & \St{$-1.3+0.1i$\\$(-0.8-0.1i)$}\\ 
		\hline
		3rd pole~$\frac{3}{2}^-$ & $(1.6+1.6i)$ & $(-13.0+14.9i)$ & $(-3.8-1.1i)$ & $(-5.4+0.1i)$ & $(3.9-8.1i)$\\ 
		\hline
		$\Delta(1930)\,\frac{5}{2}^-$ & \St{$-8.6+9.7i$\\$(-9.7+11.8i)$} & \St{$-0.1-0.6i$\\$(0.0-0.5i)$} & \St{$2.5-1.1i$\\$(3.3-0.4i)$} & \St{$1.0-0.2i$\\$(2.1+0.6i)$} & \St{$1.0+0.2i$\\$(0.3+1.1i)$}\\ 
		\hline
		$\Delta(1905)\,\frac{5}{2}^+$ & \St{$-0.2-0.1i$\\$(-0.2-0.0i)$} & \St{$-1.4+1.4i$\\$(-1.6+2.5i)$} & \St{$-0.3-0.0i$\\$(-0.0+0.2i)$} & \St{$0.3+0.5i$\\$(0.7+0.9i)$} & \St{$2.7+4.7i$\\$(5.3+7.7i)$}\\ 
		\hline
		$\Delta(1950)\,\frac{7}{2}^+$ & \St{$-6.4+1.3i$\\$(-6.5+1.3i)$} & \St{$0.2-0.6i$\\$(0.2-0.5i)$} & \St{$2.1+1.8i$\\$(2.1+1.8i)$} & \St{$4.4+5.1i$\\$(4.5+5.1i)$} & \St{$-0.3+0.5i$\\$(-0.3+0.4i)$}\\ 
		\hline
		$\Delta(2200)\,\frac{7}{2}^-$ & \St{$0.0+0.7i$\\$(0.8+2.4i)$} & \St{$3.4+0.5i$\\$(4.1+2.0i)$} & \St{$0.5-2.2i$\\$(0.9-2.9i)$} & \St{$0.8-1.9i$\\$(2.3-3.3i)$} & \St{$1.3+2.0i$\\$(4.4+3.4i)$}\\ 
		\hline
		2nd pole~$\frac{7}{2}^-$ & $(-1.2+0.3i)$ & $(-7.6+6.2i)$ & $(1.9+1.5i)$ & $(3.4+2.0i)$ & $(3.9+4.8i)$\\ 
		\hline
		$\Delta(2400)\,\frac{9}{2}^-$ & \St{$7.4-9.5i$\\$(7.4-9.5i)$} & \St{$-0.0+0.9i$\\$(-0.0+0.9i)$} & \St{$-1.9+1.0i$\\$(-2.0+0.9i)$} & \St{$0.6+1.1i$\\$(0.6+1.1i)$} & \St{$-0.0-0.2i$\\$(-0.0-0.2i)$}\\ 
	\end{tabular}
	\end{ruledtabular}
	\caption{The complex couplings of $\Delta$ states to the effective three-body channels
          (in units of $10^{-3}\sqrt{\text{MeV}}$), defined by Eq.~\eqref{Tpole}. The values
          outside (inside) the brackets are from fit A (B). $\pi\Delta$ (1,2) stand for $|J-L|=\frac{1}{2},\frac{3}{2}$, respectively; while the three sub-channels of $\rho N$ are: (1)$|J-L|=\frac{1}{2},S=\frac{1}{2}$; (2)$|J-L|=\frac{1}{2},S=\frac{3}{2}$; (3)$|J-L|=\frac{3}{2},S=\frac{3}{2}$. }
	\label{tab:threebodyDel}
\end{table*}

\clearpage
\bibliographystyle{h-physrev}
\bibliography{oN_ref}

\end{document}